\RequirePackage{snapshot}
\RequirePackage[final]{graphicx}
\documentclass[twocolumn,aps,prd,10pt,nofootinbib,nobibnotes]{revtex4-2}

\usepackage{amsmath,amssymb,amsfonts,amsthm}

\usepackage{amsmath}
\numberwithin{equation}{section}
\renewcommand{\thesection}{\arabic{section}}

\usepackage{enumerate}
\usepackage{enumitem}
\usepackage{epsfig}
\usepackage{latexsym}
\usepackage{xcolor}
\usepackage{mathrsfs}
\usepackage[colorinlistoftodos, shadow, textsize=small, obeyDraft,loadshadowlibrary]{todonotes}

\usepackage[breaklinks,colorlinks=true,citecolor=blue,linkcolor=blue,final]{hyperref}

\usepackage{parskip}
\usepackage{mciteplus}

\usepackage{comment}

\newcommand{\bite}{\begin{itemize}}
	\newcommand{\eat}{\end{itemize}}
\newcommand{\beq}{\begin{equation}}
	\newcommand{\eeq}{\end{equation}}
\newcommand{\rarrow}{\rightarrow}
\newcommand{\beqa}{\begin{align}}
	\newcommand{\eeqa}{\end{align}}
\newcommand{\barr}{\begin{array}}
	\newcommand{\earr}{\end{array}}
\newcommand{\del}{\partial}

\newcommand{\C}{\mathbb{C}}

\newcommand{\ie}{\textit{i.e.}~}
\newcommand{\eg}{\textit{e.g.}~}
\newcommand{\wrt}{\textit{w.r.t.}~}

\newcommand{\ep}{\epsilon}

\newcommand{\mb}[1]{\mathbf{#1}}
\newcommand{\mc}[1]{\mathcal{#1}}
\newcommand{\mf}[1]{\mathfrak{#1}}
\newcommand{\mbb}[1]{\mathbb{#1}}

\usepackage{bbold}

\newcommand{\bra}[1]{\langle #1 \vert}
\newcommand{\ket}[1]{\vert #1 \rangle}

\newcommand{\sltwoc}{\mathfrak{sl}(2,\mathbb{C})}

\newcommand{\onehalf}{\frac{1}{2}}

\newcommand{\Tr}{\mathrm{Tr}}

\newcommand{\diff}{\mathrm{d}}

\newcommand{\e}{\mathrm{e}}
\newcommand{\Lagr}{\mathscr{L}}
\newcommand{\PS}{\mathcal{P}}

\graphicspath{{figures/}}

\begin{document}

\preprint{This line only printed with preprint option}

\title{A Loop Quantum Gravity Inspired Action for the Bosonic String and Emergent Dimensions at Large Scales}

\author{Deepak Vaid}
\email{dvaid79@gmail.com}
\affiliation{}
\author{Luigi Teixeira de Sousa}
\email{luigi.tiraque@gmail.com}
\affiliation{Universidade Federal de São Carlos, Brasil}

\date{\today}

\begin{abstract}

	We propose a modification of the Nambu-Goto action for the bosonic string which is compatible with the existence of a minimum area at the Planck scale. The result is a phenomenological action based on the observation that LQG tells us that areas of two-surfaces are operators in quantum geometry and are bounded from below. This leads us to a string action which is similar to that of bimetric gravity. We provide formulations of the bimetric string action for both the Nambu-Goto (second order) and Polyakov (first order) formulations. We explore the classical solutions of this action and its quantization and relate it to the conventional string solutions.

	We further construct a string action in which the effect of the background geometry is described in terms of the pullback of the bulk connection, which encodes the bulk geometry, to the worldsheet. The resulting string action is in the form of a gauged sigma model, where the spacetime co-ordinates are now vectors which transform under the Poincaré group $ISO(D,1)$. This requires the introduction of an auxiliary bulk co-ordinate which has a natural interpretation as a holographic or scale direction. We discuss possible cosmological implications of such a large scale emergent dimension.
\end{abstract}

\maketitle

\tableofcontents

\listoftodos

\section{Introduction}\label{sec:intro}

``Quantum gravity'' refers to the broad enterprise dedicated to finding a complete, consistent theory in which quantum mechanics coexists peacefully with general relativity. There are two major approaches to quantum gravity which are widely recognized for their success in describing various aspects of physics at the Planck scale.

The first of these is \emph{string theory} \cite{Polchinski1998String,Polchinski2007String}, the quantum theory of a one dimensional object which is said to contain within it a complete description of all particles, forces and their interactions, including gravity. However, where precisely in the vast space of effective field theories which can possibly arise as in the low energy limit of a conjectured ``M-Theory'', our Universe with its collection of particles, forces and coupling constants, exists, is still a matter of vigorous debate.

The second approach is called \emph{loop quantum gravity} or ``LQG'' for short \cite{Rovelli1998Loop,Rovelli2015Covariant}. LQG is ostensibly meant to be a theory of ``pure quantum gravity'', that is gravity without any matter. Matter can always be ``added'' in by hand into the framework but is not present to begin with. LQG is also built upon a rigorous mathematical framework which applies the well-established Dirac procedure of quantization to canonical (ADM) Einstein gravity, and this is crucial, written in terms of Ashtekar variables which is a connection based approach as opposed to the conventional metric formalism. LQG has concrete predictions such as discreteness of physical areas \cite{Rovelli1993Area,Rovelli1994Discreteness}, and also has rigorous mathematical results which reproduce the black hole entropy formulae for Einstein gravity \cite{Rovelli1996Black}. Moreoever it allows us to construct a non-perturbative framework for quantum geometry which makes it possible to probe regimes which do not appear to be accessible via string theory.

In this paper we present a phenomenological modification of the Nambu-Goto action for the bosonic string which has the property that the worldsheet area can never be less than some fixed constant value $\Delta$ which represents the minimum allowed quantum of area predicted by LQG.

The following is the outline of the paper along with a brief summary of what is accomplished in each section. For those seeking a rapid review of the principle areas of diffence between the two theories \autoref{sec:history} provides a bird's eye view of research spanning a period of nearly sixty years.

Those already familiar with the essentials of LQG (primarily area quantization) and the question of background independence of string theory can skip ahead to \autoref{sec:modified-ng} where we being by recalling the Nambu-Goto action and suggest a simple modification which incorporates our \textbf{central dogma}, which is that \emph{there must exist a smallest possible area $\Delta$ which is physically measurable.} \footnote{In recent work \cite{Borissova2024From} a very similar expression to ours for the modified Nambu-Goto action has been arrived at by considering a version of LQG called ``area metric gravity''.} We show that this action can be expanded in terms of a small coupling constant $k = l_{pl} / l_s$ where the Planck length - the length at which quantum geometric effects become important - $l_{pl}$ is much smaller than the string scale $l_s$ and therefore $k \ll 1$. We point out that the lowest order correction to the NG action can be interpreted in the context of bimetric gravity. In \autoref{sec:bimetric-string} we expand upon this observation and suggest a new action for the ``bimetric string'', along the lines of bimetric gravity. We find both, the Nambu-Goto and the Polaykov forms of this action. Furthermore, we show that the phenomenological correction $\Delta$ introduced earlier is related to the pullback of the Kalb-Ramond field $B_{\mu \nu}$ into the worldsheet.

In \autoref{sec:inverse-metric-Polyakov} we explore the Polyakov version of the modified string action derived from the bimetric analysis and point out its consequences for the equations of motion. We also take a glance at the explicit expression for the $\Delta _{a b}$ field by considering the (area-corrected) string coupled to the Kalb-Ramond field. Furthermore, we explore an equivalence between our proposed model and the Born-Infeld model in $1 + 1$ dimensions. We conclude this section by solving the string equations of motion for a particular case of constant-coefficient to find the fundamental modes and derive the Poisson bracket algebra for the modes and constraints. In \autoref{sec:quantum-kg-string} we follow the route of covariant quantization for the solution previously found and discuss the consequences of our area-corrected model regarding the dimensionality of the background spacetime and the presence of negative-norm states (ghosts) in the physical excitations.

In \autoref{sec:emergent-dimensions} we complement Afshordi's work \cite{Afshordi2014Emergent} with insights from LQG, showing that LQG quantum geometry naturally constrains most of the free parameters of said model, leading to a more concrete theory. We also suggest a possible consequence for the parameters that remain unconstrained and a generalization of the model.

In \autoref{sec:holographic-strings} we describe how to construct an action for the bosonic string which encodes the geometry of the bulk spacetime in the language of the connection formulation of general relativity. We find that this can be done by viewing the worldsheet embedding co-ordinates $X$'s as living in the fundamental representation of a suitable gauge group $G$. Consistency between the dual roles played by the $X$'s as both co-ordinates for the background manifold \emph{and} and as objects which can be parallel transported using a connection, requires that the gauge group $G$ be the Poincaré group $ISO(D+1,1)$. In other words the string which originally was embedded in a $D+1$ dimensional background, now lives in a $D+2$ dimensional geometry which has one extra ``emergent'' dimension which has the natural interpretation as a \emph{scaling} dimension. We therefore view our construction as providing a natural framework for \emph{classical holography}, where the scale dimension is already built into the theory at the classical level instead of something which arises due to non-perturbative physics.

There are various appendices containing details of calculations and background information. In Appendix \ref{sec:modified-eom} we derive the equations of motion for the modified Nambu-Goto action. Appendix \ref{sec:polyakov-ng-relation} shows how the equations of motion extracted from the Nambu-Goto and Polakov versions of the modified string action are equivalent to each other.

\subsection{Historical Background}\label{sec:history}

Traditionally, there are two ways in which a classical theory can be quantized. These are the Hamiltonian and Lagrangian approaches. The Lagrangian or path integral approach follows the prescription first suggested by Dirac and then made concrete by Feynman. There one views the classical action associated to a given evolution as corresponding to a \emph{phase angle} which determines the complex weight of the associated evolution. This approach respects spacetime covariance since here the central object is the action which is invariant under spacetime transformations by construction. String theory is an example of such a formalism.

The Hamiltonian approach, on the other hand, involves making a choice of a spacelike surface $\Sigma_t$ and a corresponding timelike vector $t^\mu$, normal to $\Sigma_t$ (here $t$ is a continuous parameter which labels the family of surfaces). This approach, therefore, necessarily does not respect spacetime covariance since it involves making a specific choice of the foliation of the background into spacelike surfaces. Loop quantum gravity is built upon this approach to quantization.

In the decades prior to the advent of string theory a great deal of effort was put into attempting to quantize gravity via both the Lagrangian and Hamiltonian approaches. The Hamiltonian approach involves, as stated earlier, a choice of the foliation of the background geometry into a collection of spacelike surfaces $\Sigma_t$. Central to this approach was the ADM (Arnowitt-Deser-Misner) formalism which allows one to construct a Hamiltonian $H_{GR}$ for general relativity starting from the Einstein-Hilbert action $S_{EH}$. The resulting phase space is coordinatized by a configuration variable which is the 3-metric $h_{ab}$ of the ``leaves'' $\Sigma_t$ of the foliation and a momentum variable which is a function of the extrinsic curvature $k_{ab}$ of the leaves\footnote{Our notational convention is the following. Lowercase Greek letters from the middle of the alphabet $\mu, \nu, \ldots$ are spacetime indices which run from $(0 \ldots 3)$, whereas lowercase Roman letters $a,b,\ldots$ are spatial indices for quantities which live solely on the surfaces $\Sigma_t$ and take values in $(1,2,3)$}. The gravity Hamiltonian turns out to be a sum of two constraints known as the diffeomorphism (or ``momentum'') constraint $H_{diff}$ and the Hamiltonian constraint $H_{ham}$:
\begin{equation}
	H_{GR} = H_{diff} + H_{ham},
\end{equation}
and these are, in turn, functionals on the phase space of general relativity written in terms of $h_{ab}$ and $k_{ab}$. The idea then is that physical states of the theory $\ket{\Psi_{phys}} $, which can be written as functionals $\Psi[h_{ab}]$ of the 3-metric $h_{ab}$ , must be annihilated by these constraints:
\begin{equation}
	H_{GR} \ket{\Psi_{phys}} \equiv 0
\end{equation}
While this procedure is straightforward in principle, in practice it was impossible to implement in the quantum theory due to the highly complicated non-polynomial dependence of the diffeomorphism and Hamiltonian constraints on the configuration and momentum variables. This unfortunate state of affairs persisted until the late 1980s when Abhay Ashtekar \cite{Ashketar1986New,Ashtekar1987New} recast general relativity as a BF theory of a dualised $\mathfrak{sl} (2 , \C)$ connection (which becomes the self-dual part of said connection in the case of $\beta = i$)
\begin{equation} \label{eqn:self-dual-spin-connection}
	A_{\mu} ^{I J} = \omega ^{I J} _{\mu} - \frac{\beta}{2} \varepsilon ^{I J} _{\ \ K L} \omega ^{K L} _{\mu}
\end{equation}
and its conjugate momenta
\begin{equation} \label{eqn:densitized-tetrad}
	B^{I J} _{\mu \nu} = \left( \delta ^{I} _{K} \delta ^{J} _{L} + \frac{1}{2 \beta} \varepsilon ^{I J} _{\ \ K L} \right) e^{K} _{[ \mu} e^{L} _{\nu ]} ,
\end{equation}
which when restricted to a 3-dimensional hypersurface where $* (e \wedge e) = 0$ reduces to
\begin{equation} \label{eqn:Ashtekar-connection-canonical}
	A^{i} _{a} = \omega ^{i} _{a} + \beta K^{i} _{a}
\end{equation}
\begin{equation} \label{eqn:Ashtekar-E-field-canonical}
	E^{a} _{i} = \sqrt{q} e^{a} _{i} = \frac{1}{2} \varepsilon ^{a b c} \varepsilon _{i j k} e^{j} _{b} e^{k} _{c} ,
\end{equation}
with $K^{i} _{a}$ being the triad version of the extrinsic curvature $K_{a b} = \frac{1}{2} \mathcal{L}_{n} q_{a b}$ and $q$ the 3-metric. Ashtekar realized that the constraints of general relativity, when expressed in terms of these ``new variables'', simplified drastically and became polynomial functions of the tetrad and connection variables. It was quickly realized by various researchers that the resulting theory could be quantized using the same methods used to quantize Yang-Mills gauge theory \cite{Ashtekar1992Weaving,Ashtekar2004Background} and that the diffeomorphism constraint could be solved exactly in terms of so-called \emph{spin-network} states \cite{Rovelli1988Knot,Rovelli1990Loop,Ashtekar1992Gravitons}.

Following this, work by Rovelli and Smolin \cite{Rovelli1993Area,Rovelli1994Discreteness} demonstrated a remarkable feature of this theory, which came to be known as ``Loop Quantum Gravity'', was that one could construct quantum operators for geometric quantities such as areas of two-dimensional surfaces and volumes of three-dimensional regions. Moreover, these operators could be diagonalized exactly in the spin-network basis and a discrete spectrum for quantum of area and quantum of volume could be derived. This was the first time that physicists had discovered the ``atoms of space'' or the ``quanta of geometry'' in the true sense of the expression, in a theory of quantum gravity. One of the early triumphs of this formalism, which has unfortunately not received enough attention over the years by the larger high energy theory community, was the calculation of the entropy of a Schwarzschild black hole \cite{Rovelli1996Black} in a fully quantum mechanical framework and showing that it matched the expected expression $S \sim A/4$ for the Bekenstein-Hawking entropy.

One key feature of the discreteness of spacetime itself is that the Ashtekar connection $A^{i} _{a}$ \footnote{Here, the index convention followed is that lowercase $i , j , k$ refer to 3-dimensional flat internal ($\mf{su}(2)$) indices and lowercase $a , b , c$ refer to 3-dimensional curved spacetime indices} can no longer be a continuous function of spacetime, rather it becomes a \emph{distribution}, assuming a constant value on one patch of space, then another value at another patch, etc \cite{Bodendorfer-LQG,Rovelli2015Covariant}. The consequence of this non-continuous distributional nature of the connection is that the triad (or Ashtekar electric field) operator, which formally would be
\begin{equation} \label{eqn:triad-operator}
	\Hat{\widetilde{E}}{}^{a} _{i} = - 8 \pi i \hbar G \frac{\delta}{\delta A^{i} _{a}}
\end{equation}
becomes ill-defined since $A^{i} _{a}$ is non-continuous (in particular, at the edges of a polyhedra it changes abruptly), so the triad needs to be \emph{smeared} over a surface where $A^{i} _{a}$ is constant
\begin{equation} \label{eqn:triad-flux-operator}
	\Hat{E}_{i} (S) = - 8 \pi i \hbar G \int _{S} \diff S_{a}  \frac{\delta}{\delta A^{i} _{a}} ,
\end{equation}
where $\diff S_{a} = \diff \sigma ^{b} \wedge \diff \sigma ^{c} \varepsilon _{a b c}$. This naturally leads to the \emph{area operator} of said surface
\begin{equation} \label{eqn:area-operator}
	\Hat{A} (S) = - 8 \pi i l_{pl} ^2 \beta \int _{S} \sqrt{\diff S_{a} \diff S_{b} \frac{\delta}{\delta A^{i} _{a}} \frac{\delta}{\delta A_{b i}}} .
\end{equation}
Notably, since $A^{i} _{a}$ is a $\mf{su}(2)$-valued connection, in the basis which diagonalises the area operator (the \emph{spin network} basis \cite{Bodendorfer-LQG}) $\ket{\Tr (h^{(j)} _{c} (A))}$, the eigenvalues of the area operator are just proportional to \emph{angular momentum} eigenvalues
\begin{equation} \label{eqn:area-eigenvalues}
	\Hat{A} \ket{\Tr (h^{(j)} _{c} (A))} = 8 \pi l_{pl} ^2 \beta \sqrt{j (j + 1)} \ket{\Tr (h^{(j)} _{c} (A))} ,
\end{equation}
with $j$ being the representation label of $\mf{su}(2)$ along the spin network edge piercing the surface under consideration. Thus, spin network edges carry angular momenta.

One of the primary criticisms of string theory is its apparent lack of ``background independence''. This is more general than the principle of \emph{general covariance} which states that physics should be independent of the choice of co-ordinates. Background independence is the statement that any \emph{quantum} theory of gravity should provide a description of physics which is independent, not only of the choice of co-ordinates, but also of the choice of background manifold on which those co-ordinates live. Such a theory should provide a description not only of classical spacetimes, but also of fluctuating, semiclassical geometries and those in the deep quantum regime which do not have any sensible description in terms of any Riemannian geometry. On the face of it, string theory, at least in its conventional form, does not satisfy this principle. The great virtue of LQG on the other hand, is that it respects background independence by construction and proceeds to quantize gravity while ensuring that (a notion of) background independence forms the core of the quantum theory.

Some will argue that string theory is, indeed, background independent. The argument for this is as follows. When coupled to the string action, background fields are treated as external ``coupling constants'' and can therefore be subject to a renormalization group analysis. The fixed point of this analysis, remarkably enough, requires one to impose the condition that the background metric obeys Einstein's field equations (more precisely that the background is \emph{Ricci flat}. This is explained in greater detail in Appendix \ref{sec:strings-gravity}) In our opinion, however, this result is insufficient evidence for asserting that string theory is background independent. Our reasons are the following:

\begin{enumerate}[itemsep=0.5em]
	\item While the conformal invariance of the string worldsheet implies that the background is Ricci flat, this, in itself, is not very surprising. After all conformal invariance of worldsheet is equivalent to the statement of diffeomorphism invariance of the worldsheet, and in order for this to be consistent with the bulk geometry, the bulk geometry should also have a description in terms of a diffeomorphism invariant theory, the simplest example of which is Einstein gravity. However, this by itself is not sufficient to characterize string theory as being background independent. Neither do we have an understanding of how to quantize a string for which the bulk is arbitrarily curved to begin with. Nor does this tell us the microscopic mechanism of ``string condensation'' via which an arbitrary curved geometry is supposed to arise.

	\item One can ask what is the quantum mechanical description of the bulk geometry? It is often suggested that the bulk geometry corresponds to a coherent state or a graviton condensate, where the graviton are themselves excitations of the fundamental string. However, there does not appear to be an actual mathematical realization of this statement in terms of non-perturbative physics.
	\item Finally, as mentioned in the introduction, in any \emph{quantum} theory of gravity one would expect to be able to describe the ``atoms of geometry''. There is no sense in which this is realised in this picture of gravity emerging from stringy physics. Even though one can calculate higher order corrections to the graviton beta function and this provides us with the quantum corrected effective action for the bulk geometry, all the physical quantities are still defined in terms of continuous fields such as the metric. For \eg, to the best of our knowledge, there is no clear sense in which one construct a \emph{non-perturbative} state which corresponds to a superposition of geometries using the quantum theory of a bosonic string alone.
\end{enumerate}

Despite all these concerns string theory has emerged as the primary candidate for a theory of quantum gravity. \emph{There are good reasons for this}. AdS/CFT - in our opinion, the most concrete example of a \emph{non-perturbative} theory of quantum gravity - arose out of string theory \cite{Maldacena1998The-Large,Witten1998Anti,Aharony1999Large,Gubser1998Gauge}. AdS/CFT is a particular realization of quantum gravity rather than a theory of quantum gravity per se.

Anti-deSitter space describes a universe with a negative cosmological constant ($\Lambda < 0$) whereas the one we inhabit appears to have a small but positive cosmological constant ($\Lambda > 0$). Of course, nowhere in the literature of the AdS/CFT correspondence is it implied that it is the AdS bulk which corresponds to our physical universe. If anything our physical universe will turn out to live on the boundary of an anti-deSitter bulk.

AdS spaces do arise in actual physical phenomena in \emph{our} universe \cite{Bardeen1999The-Extreme,Guica2009The-Kerr/CFT}. The correspondence itself has passed numerous checks and is now considered part of the established canon. Moreoever, it has led to numerous theoretical insights. We have gained a deep understanding of the properties of real world systems referred to as ``strange metals'' \cite{Hartnoll2016Holographic,Sachdev2010Strange,Sachdev2015Bekenstein-Hawking} by studying their dual gravitational theories. The significance of quantum error correction and quantum information for a theory of quantum gravity was also first most clearly illuminated via AdS/CFT \cite{Almheiri2014Bulk,Pastawski2015Holographic}. For these, and other reasons too numerous to mention here, many theorists still regard string theory as the primary (or even the sole) candidate for a theory of quantum gravity.

Thus, we are now at an impasse. We have two theories, both built upon strong mathematical foundations, but which do not appear to be in concordance with each other. There are two possibilities. The first is that one of the two approaches is fundamentally flawed in some way and must be rejected altogether. The second is that we are missing some crucial ingredient which would allow us to find a missing link between the two theories. We feel that it worthwhile to explore the second option in greater detail, if for no other reason then to rule out the possibility of such a link between these two theories.

\section{Modified Nambu Goto Action}\label{sec:modified-ng}

Before proceeding let us restate our ``central dogma''. We take as physically well-motivated the central result of LQG that at the Planck scale well-defined geometric operators - such as those that measure the area and volume of a given region of space - exist and have discrete spectra, thus the transition from no area/volume to some area/volume is a discrete jump rather than a continuous one. We now wish to understand how to construct a worldsheet action which reflects this central dogma.

In a very real sense string theory is a quantum theory of geometry. After all, the Nambu-Goto action:
\begin{equation}\label{eqn:ng-action}
	S_{NG} = -T \int \diff ^2 x \sqrt{- \det h}
\end{equation}
is nothing more than the \emph{area} of the string worldsheet. So when we are constructing the quantum theory of the string we are constructing a theory in which the fundamental excitations are geometric in nature. What we would like to do is to modify the string action in a way which incorporates the insight from LQG that at the Planck scale there is a minimum quantum of area. One simple way in which to do this is to modify the action \eqref{eqn:ng-action} as follows:
\begin{equation} \label{eqn:ng-area-corrected}
	S_{M N G} = - T \int \diff ^2 x \sqrt{- (h + k \Delta)}.
\end{equation}
Here, $k = l_{pl} / l_s$, where $l_{pl}$ is the Planck scale, the scale at which quantum geometric effects become important; $l_s$ is the string scale where quantum geometric effects due to area quantization become small; for the time being we take $\Delta$ to be a scalar constant which corresponds to the minimum quantum of area at the Planck scale\footnote{We will see later that $\Delta$ is, in fact, a scalar density of weight 2 which can be taken to be the determinant of an antisymmetric worldsheet tensor $\Delta _{a b} (x)$ and corresponds to the square of the quantum of area at the point $x$ where the edges of the bulk spin network state pierce the WS and endow it with quanta of area as described in the previous section.}. In the limit that $l_s \gg l_{pl}$, $k \ll 1$ and the string action reduces to the conventional Nambu-Goto action with small corrections coming from quantum geometry. In this limit we can expand the integrand as follows:
\begin{equation} \label{eqn:inverse-metric-correction}
	\sqrt{- (h + k \Delta)} \sim \sqrt{- h}\left(1 - \frac{1}{2} \frac{k \Delta}{(- h)} + \mc{O}(k^2/h^2)\right).
\end{equation}
Presumably the limit $k \ll 1$ corresponds to when the area of the string worldsheet is large compared to $\Delta$, $h \gg \Delta$. Thus in this limit $ k^2 \ll 1$ and $ h^2 \gg 1$ because of which we can drop the $\mc{O}(k^2/h^2)$ corrections. The final area corrected Nambu-Goto action becomes:
\begin{equation} \label{eqn:modified-ng-action}
	S'_{NG} = -T \int \diff ^2 x \left( \sqrt{- h} - \frac{1}{2} \frac{k \Delta}{\sqrt{- h}} \right).
\end{equation}
This action has, to the best of our knowledge, not been explored in the string theory literature. However, such an inverse determinant term arises very naturally in the LQG approach and goes by the name of ``inverse triad'' corrections, in the context of the 3D volume of spatial hypersurfaces.

The action eq \eqref{eqn:modified-ng-action} has manifest background Poincaré invariance since just like the regular Nambu-Goto action since all its terms are derivatives of $X^{\mu}$ with all Lorentz indices contracted, thus are Lorentz scalars. At first glance it might seem to not be WS reparameterization invariant because of the second term, however, this is corrected by the fact that $\Delta$ is a scalar \emph{density} of weight 2, thus counteracting the inverse metric determinant. Moreover, if we look at that expression we see that it can be written as:
\begin{equation} \label{eqn:modified-ng-action-v2}
	S'_{NG}[h_{ab}] = -T \int \diff ^2 x \left( \sqrt{- h} - \frac{1}{2} {k \Delta}{\sqrt{- (h^{-1})}} \right),
\end{equation}
where $h^{-1}$ is now the determinant of the \emph{inverse} metric $h^{ab}$. As such, in the expression \eqref{eqn:modified-ng-action} we can see a duality between the first and second terms. If we replace the metric $h_{ab}$ by its inverse $h'_{ab} \equiv h^{ab}$, then the action in terms of $h'_{ab}$ is the same as \eqref{eqn:modified-ng-action}, but with the difference that the factor of $k \Delta / 2$ is mapped to $2 / (k \Delta)$:
\begin{align} \label{eqn:modified-ng-action-v3}
	S'_{NG}[h'_{ab}] & = -T' \int \diff ^2 x' \left( \sqrt{- h'} - \frac{2}{k \Delta} \sqrt{- (h')^{-1}} \right) \nonumber \\
	                 & = -T' \int \diff ^2 x' \left( \sqrt{- h'} - \onehalf {k' \Delta'}\sqrt{- (h')^{-1}} \right),
\end{align}
which has the same form as the original action \eqref{eqn:modified-ng-action}, as long as we make the replacements
\begin{subequations}
	\begin{equation} \label{eqn:T'}
		T \rightarrow T' = - \frac{1}{2} k T
	\end{equation}
	\begin{equation} \label{eqn:k-Delta}
		\frac{1}{2} {k \Delta} \rarrow \frac{1}{2} {k' \Delta'} = \frac{2} {k \Delta}
	\end{equation}
	\begin{equation} \label{eqn:d2x}
		\diff ^2 x \rightarrow \diff ^2 x' = \Delta \diff ^2 x
	\end{equation}
\end{subequations}
(or equivalently $\diff x^{a} \rightarrow \diff x' _{a} = \Delta _{a b} \diff x^{b}$). This is a duality between the physics at the Planck scale and the physics at the string scale, \ie between strong coupling ($k' \gg 1$) and weak coupling ($k \ll 1$), small quantum of area ($\Delta$) and large quantum of area ($\Delta'$) and between large string tension $T$ and small string tension $T'$.

Now, one might object and say that having an action which is the sum of two Nambu-Goto actions seems strange. However, similar actions have been extensively studied under the heading of ``bimetric gravity'', with the earliest work dating as far back as 1940 in two papers \cite{Rosen1940General,Rosen1940aGeneral} by Einstein's future collaborator Nathan Rosen (the `R' in `EPR' and `ER'). In 2010, bimetric gravity gained great popularity due to the seminal paper \cite{de-Rham2011Resummation} which showed that by introducing a second \emph{reference} metric into a model for massive gravity one can get rid of the Boulware-Deser ghost \cite{Boulware1972Can-Gravitation} which otherwise plagues theories with massive gravitons. The \emph{reference} metric introduced in the work \cite{de-Rham2011Resummation} was taken to be a flat metric. However later work \cite{Hassan2012Bimetric,Hassan2012Ghost-Free} showed that the second metric need not be flat and that an action of the following form:
\begin{align}\label{eqn:hassan-bimetric}
	S & = M_g^2 \int \diff^4 x \sqrt{-g} R^{(g)} + M_f^2 \int \diff^4 x \sqrt{-f} R^{(f)} \nonumber \\
	  & + 2 m^2 M^2_{eff} \int \diff^4 x \sqrt{-g}\sum_{n=0}^{4}\beta_n e_n (\sqrt{g^{-1} f}),
\end{align}
where $g_{\mu\nu}$ and $f_{\mu\nu}$ are \emph{arbitrary} metrics, $M_g^2, M_f^2$ are the Planck masses for the two sectors respectively. $R^{(g)}, R^{(f)}$ are the respective Ricci scalars for each sector. The third term is an interaction term which is responsible for making one of the gravitons massive while the other remains massless. $M^2_{eff}$ is an \emph{effective} Planck mass given by:
\begin{equation}\label{eqn:hassan-effective-mpl}
	M^2_{eff} = \left( \frac{1}{M^2_g} + \frac{1}{M^2_f} \right)^{-1}
\end{equation}

\section{The Bimetric String}\label{sec:bimetric-string}

It would be natural to suspect that the action \eqref{eqn:modified-ng-action-v2} might be related to bimetric gravity in the target space of the string. We can see that this is indeed the case as follows.

There are two ways that a string with two metrics might arise. One possibility is that there are two sets of embedding fields $X^\mu$ and $Y^\mu$. As a result, the induced metrics on the string worldsheet would also be different for both sets of embedding fields. However, in this case, we would not be talking about \emph{one} string but two \emph{different} strings described by the respective embedding functions. One could then proceed along the lines of \eqref{eqn:hassan-bimetric} and introduce a third term for the worldsheet action which describes a coupling between the two sets of embedding fields, of the form:
\begin{equation}\label{eqn:ws-bimetric-interaction}
	S_{int} = - T_{eff} \int \diff^2 x\, \Phi(g,h) \Psi(\mb{X}, \mb{Y}).
\end{equation}
Here $\Phi(g,h)$ is some function of the \emph{induced} metrics $g_{ab}$ and $h_{ab}$ on the two worldsheets and $\Psi(\mb{X}, \mb{Y})$ is some function of the embedding fields. The resulting action would be of the form:
\begin{equation}\label{eqn:ws-bimetric-action-v1}
	S_{BNG} = -T \int \diff^2 x \, \sqrt{-h} - T' \int \diff^2 x \, \sqrt{-g} + S_{int}.
\end{equation}
Here the subscript $BNG$ stands for ``Bimetric Nambu-Goto''.

The second possibility is the following. Rather than considering two different sets of embedding fields, and consequently two different worldsheets, we can work with one set of embeddings $X^\mu(\tau,\sigma)$ and \emph{two different} bulk metrics $G_{\mu\nu}$ and $H_{\mu\nu}$. As a result, we would now get two different induced metrics on the \emph{same} string worldsheet, given by:
\begin{subequations}\label{eqn:ws-induced-metrics}
	\begin{align}
		g_{ab} & = \frac{\partial X^\mu}{\partial x^a} \frac{\partial X^\nu}{\partial x^b} G_{\mu\nu}  \\
		h_{ab} & = \frac{\partial X^\mu}{\partial x^a} \frac{\partial X^\nu}{\partial x^b} H_{\mu\nu}.
	\end{align}
\end{subequations}
We can then write down the combined action for the string worldsheet as:
\begin{equation}\label{eqn:ws-bimetric-action-v2}
	S'_{BNG} = -T \int \diff^2 x\, \sqrt{-h} - T' \int \diff^2 x \, \sqrt{-g} + S'_{int},
\end{equation}
where now $S'_{int}[g,h,\mb{X}]$ is a term describing the interaction between the two metrics living on the \emph{same} worldsheet.

\subsection{Bimetric Polyakov Action}\label{sec:bimetric-polyakov}

The expression \eqref{eqn:ws-bimetric-action-v2} is not ideal because it does not retain any memory of the form of the bulk metrics unless these are included in the interaction term $S'_{int}$. One way to manifestly exhibit the dependence of the respective worldsheet actions on the two bulk metrics is to work with the Polaykov action.

Following the general procedure given in \cite{Callan1985Strings}, we can couple the worldsheet to a bulk metric as follows:
\begin{equation}\label{eqn:polyakov-curved}
	S_{P} = -\frac{T}{2} \int \diff^2 x \sqrt{-g} g^{ab} \partial_a X^\mu \partial_b X^\nu G_{\mu\nu},
\end{equation}
where $G_{\mu\nu}$ is now the bulk or target space metric. Using this form we can now write down the worldsheet Polyakov bimetric action for two bulk metrics:
\begin{align}\label{eqn:polyakov-bimetric}
	S_{PB} = & -\frac{T}{2} \int \diff^2 x\sqrt{-g} g^{ab} \partial_a X^\mu \partial_b X^\nu G_{\mu\nu} \nonumber     \\
	         & -\frac{T'}{2} \int \diff^2 x \sqrt{-h} h^{ab} \partial_a X^\mu \partial_b X^\nu H_{\mu\nu} + \nonumber \\
	         & + S_{int} [g , h , X].
\end{align}

\subsection{(Worldsheet) Kalb Ramond and Area Quantization} \label{sec:ws-kalb-ramond}

In the following, we will derive the equations of motion for the bimetric Polyakov action and along the way we will discover that the scalar density $\Delta(x)$ can be identified as being the determinant of (a tensor proportional to) the antisymmetric symbol $\varepsilon _{a c}$ living on the worldsheet. The worldsheet antisymmetric symbol can itself be understood as arising from the pullback of a bulk antisymmetric tensor. Now, in string theory, there is already a very important antisymmetric background tensor: the Kalb-Ramond field.

Following the usual procedure such as the one found in \cite{Tong2009Lectures}, we find two copies of the string energy-momentum tensor:
\begin{equation} \label{eqn:variation-wrt-g}
	\frac{\delta S_{P B}}{\delta g^{c d}} = 0
\end{equation}
\begin{multline} \label{eqn:g-em-tensor}
	T^{(G)} _{c d} := G_{\mu \nu} \bigg( \partial _{c} X^{\mu} \partial _{d} X^{\nu} - \\
	- \frac{1}{2} g_{c d} g^{a b} \partial _{a} X^{\mu} \partial _{b} X^{\nu} \bigg) = \frac{2}{T} \frac{\delta S_{int}}{\delta g^{c d}} ,
\end{multline}
and completely analogous for the $H_{\mu \nu}$ metric,
\begin{equation} \label{eqn:variation-wrt-h}
	\frac{\delta S_{P B}}{\delta h^{c d}} = 0
\end{equation}
\begin{multline} \label{eqn:h-em-tensor}
	T^{(H)} _{c d} := H_{\mu \nu} \bigg( \partial _{c} X^{\mu} \partial _{d} X^{\mu} - \\
	- \frac{1}{2} h_{c d} h^{a b} \partial _{a} X^{\mu} \partial _{b} X^{\nu} \bigg) = \frac{2}{T'} \frac{\delta S_{int}}{\delta h^{c d}} .
\end{multline}
If we assume no interaction term $S_{int} = 0$, we get the expected expressions for the auxiliary metrics $g_{c d}$ and $h_{c d}$ as in
\begin{equation} \label{eqn:g-expression}
	g_{c d} = 2 f^{(G)} \partial _{c} X^{\mu} \partial _{d} X^{\nu} G_{\mu \nu} ,
\end{equation}
where the function $f^{(G)}$ is given by
\begin{equation} \label{eqn:g's-f-function}
	\frac{1}{f^{(G)}} = g^{a b} \partial _{a} X^{\mu} \partial _{b} X^{\nu} G_{\mu \nu}
\end{equation}
and for $h_{c d}$
\begin{equation} \label{eqn:h-expression}
	h_{c d} = 2 f^{(H)} \partial _{c} X^{\mu} \partial _{d} X^{\nu} H_{\mu \nu} ,
\end{equation}
with $f^{(H)}$ given by
\begin{equation} \label{eqn:h's-f-function}
	\frac{1}{f^{(H)}} = h^{a b} \partial _{a} X^{\mu} \partial _{b} X^{\nu} H_{\mu \nu} .
\end{equation}
If the auxiliary metrics $g_{c d}$ and $h_{c d}$ coincide with the pullback metrics from $G_{\mu \nu}$ and $H_{\mu \nu}$, the action with no interaction reduces to
\begin{equation} \label{eqn:double-ng-action}
	S = - T \int \diff ^2 x \sqrt{- g} - T' \int \diff ^2 x \sqrt{- h} ,
\end{equation}
and from here we note that if $T' = - T k / 2$ and $h = \Delta ^2 / g$, we recover the Nambu-Goto area-corrected action \eqref{eqn:modified-ng-action}
\begin{equation} \label{eqn:area-corrected-ng-from-bimetric}
	S = - T \int \diff ^2 x \left( \sqrt{- g} - \frac{1}{2} \frac{k \Delta}{\sqrt{- g}} \right) .
\end{equation}
The condition $h = \Delta ^2 / g$ can be achieved by
\begin{equation} \label{eqn:h=SgS}
	h_{a b} = \Delta _{a c} \Delta _{b d} g^{c d} .
\end{equation}
By writing the inverse metric $g^{a b}$ in terms of Levi-Civita symbols $g^{a b} = \varepsilon ^{a c} \varepsilon ^{b d} g_{c d} / 2 g$, this implies that
\begin{multline} \label{eqn:SgS-implication}
	\left( \delta _{a} ^{c'} \delta _{b} ^{d'} H_{\mu \nu} - \frac{1}{2 g} \Delta _{a c} \varepsilon ^{c c'} \Delta _{b d} \varepsilon ^{d d'} G_{\mu \nu} \right) \times \\
	\times \del _{c'} X^{\mu} \del _{d'} X^{\nu} = 0 ,
\end{multline}
which can be achieved if the expression in parenthesis vanishes. For this end, the most natural possibility is
\begin{equation} \label{eqn:S=sigma_y}
	\Delta _{a c} = \sqrt{\Delta} \varepsilon _{a c}
\end{equation}
since we must have that $\det (\Delta _{a b}) = \Delta$ and $\Delta _{a c} \varepsilon ^{c c'} \sim \delta ^{c'} _{a}$, and also if the background metrics are related by
\begin{equation} \label{eqn:H=1/(2g)G}
	H_{\mu \nu} = \frac{\Delta}{2 g} G_{\mu \nu} .
\end{equation}
This makes sense since $\Delta _{a b}$ is related to area, so it should be a 2-form. Now, since the WS metric $g_{a b}$ is taken as the pullback of the bulk metric $g_{a b} = \del _{a} X^{\mu} \del _{b} X^{\nu} G_{\mu \nu}$, which is part of the closed string spin-2 spectrum, then it would make sense for the WS area 2-form $\Delta _{a b}$ to be (at least related to) the pullback of the Kalb-Ramond field, which is the other half of the closed string rank 2 tensor field excitations,
\begin{equation} \label{eqn:Delta-B-relation}
	\Delta _{a b} \sim \del _{a} X^{\mu} \del _{b} X^{\nu} B_{\mu \nu} .
\end{equation}

If we impose the conditions \eqref{eqn:h=SgS}, \eqref{eqn:S=sigma_y} and \eqref{eqn:H=1/(2g)G} plus $T' = - T k / 2$ in the interactionless bimetric Polyakov action \eqref{eqn:polyakov-bimetric}, we obtain the Polyakov version of the area-corrected action \eqref{eqn:modified-ng-action}:
\begin{multline} \label{eqn:inverse-metric-Polyakov}
	S_{I A P} = - \frac{T}{2} \int \diff ^2 x \left( \sqrt{- g} - \frac{k \Delta}{2 \sqrt{- g}} \right) \times \\ \times g^{a b} \partial _{a} X^{\mu} \partial _{b} X^{\nu} G_{\mu \nu} .
\end{multline}

\section{Inverse Metric Polyakov} \label{sec:inverse-metric-Polyakov}

In this section, we explore the action \eqref{eqn:inverse-metric-Polyakov} derived previously. This action is even more manifestly background Poincaré invariant, since now the $X^{\mu}$ dependence is explicit with this being a first-order action with $g_{a b}$ and $X^{\mu}$ treated as independent variables. This action at first glance does not have WS Weyl invariance since the second term has inverse determinant multiplied by inverse metric, however, if one wishes to restore Weyl symmetry one just needs to require that under a Weyl transformation, the $\Delta _{a b}$ field also transforms like the metric \ie
\begin{subequations}
	\begin{align}
		g_{a b}       & \rightarrow \Lambda (x) g_{a b} \label{eqn:g-Weyl-transform}             \\
		\Delta _{a b} & \rightarrow \Lambda (x) \Delta _{a b} , \label{eqn:Delta-Weyl-transform}
	\end{align}
\end{subequations}
however, we will not work under this assumption, since we expect that Weyl symmetry is broken near the Planck scale due to the discreteness of space-time.

The equations of motion for the world-sheet metric are completely unchanged by this correction,
\begin{equation} \label{eqn:vary-wrt-g}
	\frac{\delta S_{I A P}}{\delta g^{c d}} = 0
\end{equation}
which leads to the stress-energy tensor for the worldsheet:
\begin{multline} \label{eqn:inverse-area-em-tensor}
	T_{c d} := \partial _{c} X^{\mu} \partial _{d} X^{\nu} G_{\mu \nu} - \\ - \frac{1}{2} g_{c d} g^{a b} \partial _{a} X^{\mu} \partial _{b} X^{\nu} G_{\mu \nu} = 0 ,
\end{multline}
from which we can extract the metric $g_{c d}$ as
\begin{equation} \label{eqn:ws-metric}
	g_{c d} = 2 f \partial _{c} X^{\mu} \partial _{d} X^{\nu} G_{\mu \nu} ,
\end{equation}
where the function $f$ is given by
\begin{equation} \label{eqn:inverse-area-f}
	\frac{1}{f} = g^{a b} \partial _{a} X^{\mu} \partial _{b} X^{\nu} G_{\mu \nu}
\end{equation}
just like the regular ST literature, thus the constraints are unchanged by our correction.

As for variation \wrt the embbeding fields $X^{\lambda}$, we have
\begin{equation} \label{eqn:vary-wrt-X}
	\frac{\delta S_{I A P}}{\delta X^{\lambda}} = 0
\end{equation}
\begin{multline} \label{eqn:X-eom-inverse-area}
	F^{-} \sqrt{- g} g^{a b} \partial _{a} \partial _{b} X^{\nu} G_{\lambda \nu} + \partial _{a} \left( F^{-} \sqrt{- g} g^{a b} \right) \partial _{b} X^{\nu} G_{\lambda \nu} = \\ = \frac{1}{2} F^{-} \sqrt{- g} g^{a b} \partial _{a} X^{\mu} \partial _{b} X^{\nu} \partial _{\lambda} G_{\mu \nu} ,
\end{multline}
where
\begin{equation} \label{eqn:F-minus-def}
	F^{-} \equiv 1 - \frac{k \Delta}{2 (- g)} .
\end{equation}
Since all 2-dimensional metrics are conformally flat \cite{Schottenloher2008A-Mathematical}, we use conformal gauge $g_{a b} = \e ^{\phi (x)} \eta _{a b}$ and assume flat background $G_{\mu \nu} = \eta _{\mu \nu}$, to simplify eq (\ref{eqn:X-eom-inverse-area}) into
\begin{equation} \label{eqn:X-eom-simplified}
	\eta ^{a b} \partial _{a} \partial _{b} X^{\mu} + \frac{1}{F^{-}} \eta ^{a b} \partial _{a} F^{-} \partial _{b} X^{\mu} = 0 .
\end{equation}
By expanding the definition of $F^{-}$ we get
\begin{equation} \label{eqn:coupling-X-phi}
	\eta ^{a b} \partial _{a} \partial _{b} X^{\mu} + \eta ^{a b} \frac{\left( 2 \del _{a} \phi - \del _{a} \ln (\Delta) \right)}{\dfrac{2 \e ^{2 \phi}}{k \Delta} - 1} \del _{b} X^{\mu} = 0 ,
\end{equation}
revealing a wave equation sourced by a coupling between the WS conformal factor, quantum of area field $\Delta$ and the embedding fields much akin to the dilaton field \cite{Balthazar-t-backgrounds}. We'll come back to this in \autoref{sec:emergent-dimensions}.

\subsection{Worldsheet Kalb-Ramond field and Quantum of Area} \label{sec:WS-Kalb-Ramond-QoA}

If one were to calculate the equations of motion for the quantum of area field $\Delta _{a b}$ in the actions \eqref{eqn:modified-ng-action} and \eqref{eqn:inverse-metric-Polyakov} explored before, one would end up concluding that either $\Delta _{a b} = 0$ or it's determinant $\Delta = 0$ vanish. This happens because, as we explored in section \ref{sec:ws-kalb-ramond}, $\Delta _{a b}$ is actually related to the pullback of the Kalb-Ramond field $B_{\mu \nu}$, but those actions were actions for the free string, not coupled to said field. The more complete picture would be thus to consider the (area-corrected) string now coupled to the Kalb-Ramond field \cite{Tong2009Lectures}
\begin{multline} \label{eqn:area-corrected-K-R-action}
	S = - \frac{1}{4 \pi \alpha '} \int \diff ^2 x \left( \sqrt{- g} - \frac{1}{2} \frac{k \Delta}{\sqrt{- g}} \right) \times \\
	\times \Big( g^{a b} \del _{a} X^{\mu} \del _{b} X^{\nu} G_{\mu \nu} - \varepsilon ^{a b} \del _{a} X^{\mu} \del _{b} X^{\nu} B_{\mu \nu} \Big) ,
\end{multline}
where  we can rewrite $\varepsilon ^{a b}$ in terms of $\Delta ^{a b}$ as
\begin{multline} \label{eqn:area-corrected-K-R-action-v2}
	S = - \frac{1}{4 \pi \alpha '} \int \diff ^2 x \left( \sqrt{- g} - \frac{1}{2} \frac{k \Delta}{\sqrt{- g}} \right) \times \\
	\times \Big( g^{a b} \del _{a} X^{\mu} \del _{b} X^{\nu} G_{\mu \nu} - \\
	- \sqrt{\Delta} \Delta ^{a b} \del _{a} X^{\mu} \del _{b} X^{\nu} B_{\mu \nu} \Big) .
\end{multline}
By varying this action \wrt $\Delta ^{c d}$ we get
\begin{equation} \label{eqn:Delta-pullback}
	\Delta _{c d} = \widetilde{f} \del _{c} X^{\mu} \del _{d} X^{\nu} B_{\mu \nu}
\end{equation}
with $\widetilde{f}$ given by
\begin{multline} \label{eqn:1/f-tilde}
	\frac{1}{\widetilde{f}} = \frac{1}{g + \dfrac{k \Delta}{2}} \Bigg( \frac{k \sqrt{\Delta}}{2} g^{a b} \del _{a} X^{\mu} \del _{b} X^{\nu} G_{\mu \nu} - \\
	- \left( g + \frac{3 k \Delta}{4} \right) \Delta ^{a b} \del _{a} X^{\mu} \del _{b} X^{\nu} B_{\mu \nu} \Bigg) ,
\end{multline}
where if we take $g_{a b} = \del _{a} X^{\mu} \del _{b} X^{\nu} G_{\mu \nu}$ and $\del _{a} X^{\mu} \del _{b} X^{\nu} B_{\mu \nu} = \Delta _{a b} / \widetilde{f}$ this simplifies to
\begin{equation} \label{eqn:f-tilde}
	\widetilde{f} = - \frac{g + k \Delta}{k \sqrt{\Delta}} ,
\end{equation}
from which it is easy to see that the pullback of the Kalb-Ramond field $b_{a b} := \del _{a} X^{\mu} \del _{b} X^{\nu} B_{\mu \nu}$ is given by
\begin{equation} \label{eqn:WS-b-field}
	b_{a b} = - \frac{1}{\dfrac{g}{k \Delta} + 1} \varepsilon _{a b} ,
\end{equation}
or equivalently that the relation between $\Delta _{a b}$ and $b_{a b}$ is
\begin{equation} \label{eqn:Delta-b-relation}
	\Delta _{a b} = \sqrt{\Delta} \varepsilon _{a b} = - \sqrt{\Delta} \left( \frac{g}{k \Delta} + 1 \right) b_{a b} .
\end{equation}
We thus see that the pullback of the Kalb-Ramond field is actually a tensor \emph{density} of weight $- 1$, since the coefficient in eq \eqref{eqn:WS-b-field} is a true scalar and the Levi-Civita symbol is itself a density.

\subsection{Area-Corrected Nambu-Goto Action and Born-Infeld Electrodynamics} \label{sec:area-corrected-BI}

The Nambu-Goto area-corrected action before approximations \eqref{eqn:ng-area-corrected} has a structure strikingly similar to the Born-Infeld model \cite{Bergshoeff1987BornInfeld,Leigh1989Dirac-Born-Infeld,Metsaev1987Borninfeld,Tseytlin1996Self-Duality}, which arises in the string theory description of gauge fields living on D-branes:
\begin{equation} \label{eqn:BI-action}
	S_{B I} = - T \int \diff ^p x \sqrt{- \det (\gamma _{a b} + 2 \pi \alpha ' F_{a b})} ,
\end{equation}
with $T$ being the tension of the $p$-brane, $\gamma _{a b}$ is the pullback metric of said brane, $F_{a b}$ is the pullback of the field strength of the gauge field and $2 \pi \alpha '$ the inverse tension of the string. Remarkably, if one considers a $(1 + 1)$-dimensional Born-Infeld action with $2 \pi \alpha ' F_{a b} = \sqrt{k} \Delta _{a b}$, it can be shown that
\begin{equation} \label{eqn:area-corrected-NG-BI-equivalence}
	\det (h_{a b} + \sqrt{k} \Delta _{a b}) \equiv h + k \Delta ,
\end{equation}
thus the action \eqref{eqn:ng-area-corrected} can be recast as a Born-Infeld model for the string with $k$ playing the role of inverse tension squared and $\Delta _{a b}$ the role of the pullback of a field strength. Since we explored the relation between $\Delta _{a b}$ and the Kalb-Ramond field $B_{\mu \nu}$ in \autoref{sec:WS-Kalb-Ramond-QoA}, this could mean that the Kalb-Ramond 2-form is exact, $B = \diff A$, and thus its exterior derivative, which gives the torsion of the background spacetime \cite{Tong2009Lectures}, vanishes $\diff B = \diff (\diff A) = 0$. Also, we have a direct relation between string tension and scale
\begin{equation} \label{eqn:k-T}
	\frac{l_{pl}}{l_s} = k \sim (2 \pi \alpha ')^2 = \frac{1}{T^2} .
\end{equation}

\subsection{Simplifying The Equations of Motion} \label{sec:solving-eoms-inverse-area}

Equation (\ref{eqn:X-eom-simplified}) seems a bit odd, but we can polish it by choosing a plane-wave ansatz for the embedding fields $X^{\mu} = X^{\mu} _{0} \e ^{-i (E \tau - p \sigma)}$ and using it to fix a form for the conformal factor $\e ^{\phi (x)}$ appearing in $F^{-}$. By using the ansatz in eq (\ref{eqn:X-eom-simplified}), it becomes
\begin{equation}
	(E^{2} - p^{2}) X^{\mu} + i (E \partial _{\tau} \ln (F^{-}) - p \partial _{\sigma} \ln (F^{-})) X^{\mu} = 0 ,
\end{equation}
from which we can use the freedom in the conformal factor to fix the second term as $- \mu ^2 X^{\mu}$, with $- \mu ^2 = p^2 - E^2$ on-shell and this, in turn, implies that the derivatives of $\ln (F^{-})$ are given by
\begin{equation} \label{eqn:derivatives-F-minus}
	\partial _{\tau} \ln (F^{-}) = i E , \ \partial _{\sigma} \ln (F^{-}) = i p ,
\end{equation}
thus we have
\begin{equation} \label{eqn:ln-F}
	\ln (F^{-}) = i (E \tau + p \sigma) + a , \ a \in \mathbb{C}
\end{equation}

\begin{equation} \label{eqn:F-minus-exp}
	F^{-} = A \e ^{i (E \tau + p \sigma)} , \ A \in \C
\end{equation}
\begin{equation} \label{eqn:phi}
	\e ^{\phi (x)} = \sqrt{\frac{k \Delta}{2 \left( 1 - A \e ^{i (E \tau + p \sigma)} \right)}} .
\end{equation}
It may seem problematic to have a complex conformal factor, but the complex structure comes from our ansatz for the embedding fields, and when we choose either the real or imaginary part of the latter it should also pick the respective component of $\e ^{\phi}$. With this choice, we thus turned eq (\ref{eqn:X-eom-simplified}) into a Klein-Gordon equation
\begin{equation} \label{eqn:X-Klein-Gordon-eqn}
	\left( \partial _{\sigma} ^2 - \partial _{\tau} ^2 \right) X^{\mu} - \mu ^2 X^{\mu} = 0 .
\end{equation}
The imposition that
\begin{equation} \label{eqn:mu-constraint-1}
	i (E \partial _{\tau} \ln (F^{-}) - p \partial _{\sigma} \ln (F^{-})) = - \mu ^2
\end{equation}
can be used to find the explicit dependence of $\mu ^2$ on the quantum geometry parameters $k \Delta$ as
\begin{multline} \label{eqn:mu-constraint-final}
	\mu ^2 = \frac{i}{\dfrac{2 \e ^{2 \phi}}{k \Delta} - 1} \big( p \del _{\sigma} \ln \left( \e^{2 \phi} / \Delta \right) - \\
	- E \del _{\tau} \ln \left( \e^{2 \phi} / \Delta \right) \big) .
\end{multline}

\subsection{Solving The String Klein-Gordon Equation}

In the following, we focus on the analysis of the closed string unless otherwise stated.

From our analysis of the inverse-area corrected Polyakov action, we got $D$ finite-space one-dimensional Klein-Gordon equations (\ref{eqn:X-Klein-Gordon-eqn}), with constraints given by eq \eqref{eqn:inverse-area-em-tensor}. The constraints are nothing new, they're the same as the ones of regular String Theory. What's new in our analysis is the $\mu ^2$ term, which we take as being constant for a first analysis, making it so the solution is given by Fourier transform
\begin{multline} \label{eqn:KG-Polyakov-solution-initial}
	X^{\mu} (\tau , \sigma) = \sqrt{\frac{\alpha '}{2}} \sum _{p \in \mathbb{Z}} \frac{1}{2 E_{p}} \bigg( \alpha ^{\mu} _{p} \e ^{i (E_{p} \tau - p \sigma)} + \\
	+ \beta ^{\mu} _{p} \e ^{- i (E_{p} \tau - p \sigma)} \bigg) ,
\end{multline}
with $E_{p} = \sqrt{p^2 + \mu ^2}$. From here on, following an analogous procedure to the one found in \cite{Tong2009Lectures}, reality of $X^{\mu}$ implies $(X^{\mu})^{*} = X^{\mu}$, from which we get
\begin{equation} \label{eqn:reality-consequence}
	\beta ^{\mu} _{n} = (\alpha ^{\mu} _{n})^{*} .
\end{equation}

When looking at the constraints, we see that we can't decouple the exponentials in $\tau$ from any of the sums, thus we get two-indices \textquotedblleft Virasoro modes"
\begin{subequations}
	\begin{align}
		L_{n , p} := \alpha ^{\mu} _{n} \alpha ^{\nu} _{p} \eta _{\mu \nu}                                                         & = 0 , \ \forall \ n , p \in \mathbb{Z} , \label{eqn:"virasoro"-modes-1} \\
		\widetilde{L}_{n , p} := \alpha ^{\mu} _{n} (\alpha ^{\nu} _{p})^{*} \eta _{\mu \nu}                                       & = 0 , \ n \neq p , \label{eqn:"virasoro"-modes-2}                       \\
		\sum _{n} \left( \left( 1 + \frac{n}{E_{n}} \right) ^2 \alpha ^{\mu} _{n} (\alpha ^{\nu} _{n})^{*} \eta _{\mu \nu} \right) & = 0 \label{eqn:"virasoro"-modes-sum} .
	\end{align}
\end{subequations}
Constraint \eqref{eqn:"virasoro"-modes-2} implies that we only have $D$ independent expansion modes for each field $X^{\mu}$, which for consistency as we will see later means $\alpha ^{\mu} _{- p} = \alpha ^{\mu} _{p} , \ p \leq d = D - 1$ and $\alpha ^{\mu} _{n} = 0 , \ \forall \ |n| > d$, so we have a truncated solution. This is not surprising, since because space-time is discrete we expect wavelengths smaller than the granular structure to be ruled out.

Before proceeding, we have one remark: the 0-th term of the solution reads
\begin{equation} \label{eqn:X-sol-0-term}
	\sqrt{\frac{\alpha '}{2}} \frac{1}{2 \mu} \left( \alpha ^{\mu} _{0} \e ^{- i \mu \tau} + (\alpha ^{\mu} _{0})^{*} \e ^{i \mu \tau} \right) ,
\end{equation}
which is clearly divergent in the $\mu \rightarrow 0$ limit unless $\mathfrak{Re}(\alpha ^{\mu} _{0}) \sim \mu$. Moreover, if we want to recover the classic String Theory expression in this regime, it is not difficult to see that we require
\begin{equation} \label{eqn:a-p-relation}
	\alpha ^{\mu} _{0} = \sqrt{\frac{2}{\alpha '}} \mu x^{\mu} + i \frac{1}{2} \sqrt{\frac{\alpha '}{2}} p^{\mu} .
\end{equation}
This in actuality will differ from the normal ST literature by a factor of $1 / 2$ in the momentum term, but this is needed for the Poisson brackets to give the expected results.

Looking closer at our constraints \textquotedblleft modes", in particular the $n = p$ condition, we get the mass of the string in terms of its vibrational modes and center of mass position
\begin{multline} \label{eqn:string-rest-mass-classical-1}
	M^2 = - p^{\mu} p_{\mu} = \frac{4}{\alpha '} \Bigg( \sum _{n \neq 0} \left( \left( 1 + \frac{n}{E_{n}} \right) ^2 \widetilde{L}_{n , n} \right) - \\
	- \frac{2 \mu ^2}{\alpha '} x^{\mu} x_{\mu} \Bigg) =
\end{multline}
\begin{multline} \label{eqn:string-rest-mass-classical-2}
	= \frac{4}{\alpha '} \Bigg( \sum _{n \neq 0} \left( \left( 1 + \frac{n}{E_{n}} \right) ^2 \alpha ^{\mu} _{n} (\alpha ^{\nu} _{n})^{*} \eta _{\mu \nu} \right) - \\
	- \frac{2 \mu ^2}{\alpha '} x^{\mu} x_{\mu} \Bigg) .
\end{multline}

\subsection{The Expansion Modes Algebra}

We start by calculating the canonical momentum density conjugate to the fields $X^{\mu} (x)$, $\Pi ^{\nu} (x)$, given by
\begin{multline} \label{eqn:canonical-momentum-density}
	\Pi ^{\nu} (x) = F^{-} \frac{i}{8 \pi \alpha '} \sum _{n} \bigg( \alpha ^{\nu} _{n} \e ^{i (E_{n} \tau - n \sigma)} - \\
	- (\alpha ^{\nu} _{n})^{*} \e ^{- i (E_{n} \tau - n \sigma)} \bigg)
\end{multline}
where
\begin{multline} \label{eqn:F-minus-full-sum}
	F^{-} = 1 - \frac{k \Delta}{2 \e ^{2 \phi (x)}} = \\
	= \sum _{l} \left( A_{l} \e ^{i (E_{l} \tau + l \sigma)} + (A_{l})^{*} \e ^{- i (E_{l} \tau + l \sigma)} \right) ,
\end{multline}
and then proceed to use the equal-time Poisson bracket relation at $\tau = 0$
\begin{equation} \label{eqn:canonical-bracket-relation}
	\left\{ X^{\mu} (0 , \sigma) , \Pi ^{\nu} (0 , \sigma ') \right\} = \delta (\sigma - \sigma ') \eta ^{\mu \nu}
\end{equation}
to calculate the bracket relations between the expansion modes $\alpha ^{\mu} _{n}$ and $(\alpha ^{\nu} _{p})^{*}$. After using the freedom we still have in choosing the expansion modes of $F^{-}$ to gauge fix it to satisfy $\mathfrak{Re} (A_{0}) = - 2 \alpha '$ and realizing that $\alpha ^{\mu} _{- n} = \alpha ^{\mu} _{n}$, we get the expected
\begin{subequations}
	\begin{align}
		\left\{ \alpha ^{\mu} _{n} , \alpha ^{\nu} _{p} \right\}       & = 0 = \left\{ (\alpha ^{\mu} _{n})^{*} , (\alpha ^{\nu} _{p})^{*} \right\} \label{eqn:expansion-modes-bracket-0} \\
		\left\{ \alpha ^{\mu} _{n} , (\alpha ^{\nu} _{p})^{*} \right\} & = - i E_{n} \delta _{|n| , |p|} \eta ^{\mu \nu} , \label{eqn:expansion-modes-bracket-1}
	\end{align}
\end{subequations}
from which we can normalize the expansion modes to get harmonic oscillator modes and bracket relations
\begin{subequations}
	\begin{align}
		a^{\mu} _{n}       & := \frac{1}{\sqrt{E_{n}}} \alpha ^{\mu} _{n} \label{eqn:normalized-oscillator-modes}         \\
		(a^{\mu} _{n})^{*} & := \frac{1}{\sqrt{E_{n}}} (\alpha ^{\mu} _{n})^{*} \label{eqn:normalized-oscillator-modes-*}
	\end{align}
\end{subequations}
\begin{equation} \label{eqn:harmonic-bracket-relations}
	\left\{ a^{\mu} _{n} , (a^{\nu} _{p})^{*} \right\} = - i \delta _{|n| , |p|} \eta ^{\mu \nu} .
\end{equation}
The bracket relations for $\alpha ^{\mu} _{0}$ can be used to get the expected bracket relations between $x^{\mu}$ and $p^{\nu}$
\begin{align}
	\left\{ x^{\mu} , x^{\nu} \right\} & = 0 = \left\{ p^{\mu} , p^{\nu} \right\} \label{eqn:x-x-p-p-brackets} \\
	\left\{ x^{\mu} , p^{\nu} \right\} & = \eta ^{\mu \nu} . \label{eqn:x-p-bracket-relations}
\end{align}

\subsection{The \textquotedblleft Virasoro-Klein-Gordon" Algebra}

Having the bracket relations between the vibrational modes of the string, we can now calculate the constraints algebra, which we shall refer to as the Virasoro-Klein-Gordon Algebra. For simplicity, since $E_{n} = \sqrt{n^2 + \mu ^2} > 0$, we shall work with the normalized harmonic oscillator modes $a^{\mu} _{n}$ instead of the expansion modes $\alpha ^{\mu} _{n}$. Then, the algebra in terms of said normalized modes is given by
\begin{subequations}
	\begin{equation} \label{eqn:L-brackets-0}
		\left\{ L_{n , m} , L_{k , p} \right\} = 0
	\end{equation}
	\begin{equation} \label{eqn:L-brackets-1}
		\left\{ L_{n , m} , \widetilde{L}_{k , p} \right\} = - i \left( L_{m , k} \delta _{|n| , |p|} + L_{n , k} \delta _{|m| , |p|} \right)
	\end{equation}
	\begin{equation} \label{eqn:L-brackets-2}
		\left\{ \widetilde{L}_{n , m} , \widetilde{L}_{k , p} \right\} = - i \left( \widetilde{L}_{k , m} \delta _{|n| , |p|} - \widetilde{L}_{n , p} \delta _{|m| , |k|} \right) .
	\end{equation}
\end{subequations}

\section{The Quantum Klein-Gordon String} \label{sec:quantum-kg-string}

We now seek to quantize the Klein-Gordon string by means of covariant quantization, in a manner much similar to those found in \cite{Tong2009Lectures,Green2012Superstring}. The string coordinates $X^{\mu}$ and it's conjugate momentum $\Pi ^{\nu}$ are turned into operators $\Hat{X}^{\mu}$ and $\Hat{\Pi}^{\nu}$, with the Poisson brackets $\{ . , . \}$ being replaced by commutators $\{ . , . \} \rightarrow - i [ . , . ]$:
\begin{subequations}
	\begin{align}
		\left[ \Hat{X}^{\mu} (\tau , \sigma) , \Hat{\Pi}^{\nu} (\tau , \sigma ') \right] & = i \delta (\sigma - \sigma ') \eta ^{\mu \nu} \label{eqn:X-P-commutator}                                              \\
		\left[ \Hat{X}^{\mu} , \Hat{X}^{\nu} \right]                                     & = 0 = \left[ \Hat{\Pi}^{\mu} , \Hat{\Pi}^{\nu} \right] \label{eqn:X-X-P-P-commutators}                                 \\
		\left[ \Hat{a}^{\mu} _{n} , \Hat{a}^{\nu} _{p} \right]                           & = 0 = \left[ (\Hat{a}^{\mu} _{n})^{\dagger} , (\Hat{a}^{\nu} _{p})^{\dagger} \right] \label{eqn:a-a-a*-a*-commutators} \\
		\left[ \Hat{a}^{\mu} _{n} , (\Hat{a}^{\nu} _{p})^{\dagger} \right]               & = \delta _{|n| , |p|} \eta ^{\mu \nu} \label{eqn:a-a*-commutator}                                                      \\
		\left[ \Hat{x}^{\mu} , \Hat{x}^{\nu} \right]                                     & = 0 = \left[ \Hat{p}^{\mu} , \Hat{p}^{\nu} \right] \label{eqn:x-x-p-p-commutators}                                     \\
		\left[ \Hat{x}^{\mu} , \Hat{p}^{\nu} \right]                                     & = i \eta ^{\mu \nu} \label{eqn:x-p-commutator}
	\end{align}
\end{subequations}

As usual, we define the vacuum state of the string to obey
\begin{equation} \label{eqn:annihilation-vacuum}
	\Hat{a}^{\mu} _{p} \ket{0} = 0 , \ \mathrm{for} \ p \neq 0 .
\end{equation}
For $p = 0$ we have center of mass position and momentum, so the vacuum has extra structure obeying
\begin{subequations}
	\begin{align}
		\Hat{x}^{\mu} \ket{0 ; x} & = x^{\mu} \ket{0 ; x} \label{eqn:x-vacuum-x}                       \\
		\Hat{p}_{\mu} \ket{0 ; x} & = - i \frac{\del}{\del x^{\mu}} \ket{0 ; x} \label{eqn:p-vacuum-x}
	\end{align}
\end{subequations}
in position representation $\ket{0 ; x} \equiv \ket{0} \otimes \psi (x)$ and
\begin{subequations}
	\begin{align}
		\Hat{x}^{\mu} \ket{0 ; p} & = - i \frac{\del}{\del p_{\mu}} \ket{0 ; p} \label{eqn:x-vacuum-p} \\
		\Hat{p}_{\mu} \ket{0 ; p} & = p_{\mu} \ket{0 ; p} \label{eqn:p-vacuum-p}
	\end{align}
\end{subequations}
in momentum representation $\ket{0 ; p} \equiv \ket{0} \otimes \psi (p)$.

Our Fock space is built from the vacuum state $\ket{0}$ by operating with a sequence of creation operators
\begin{equation} \label{eqn:general-state}
	\ket{\psi} = \prod _{i \neq 0} ((\Hat{a}^{\mu _{i}} _{i})^{\dagger})^{n_{i}} \ket{0} .
\end{equation}

As usual, the appearance of the Minkowski metric in the non-zero commutator
\begin{equation} \label{eqn:ghosts}
	\Big[ \Hat{a}^{\mu} _{p} , (\Hat{a}^{\nu} _{k})^{\dagger} \Big] = \eta ^{\mu \nu} \delta _{|p| , |k|} ,
\end{equation}
brings ghosts into the theory which must be dealt with.

When translating the constraints as quantum operators, we impose them strongly, requiring that they have vanishing matrix elements when sandwiched by physical states $\psi$ and $\phi$ since said constraints are not self-adjoint:
\begin{subequations}
	\begin{equation} \label{eqn:L-constraint-quantum-1}
		\bra{\phi} \Hat{L}_{p , k} \ket{\psi} = 0 \, \,
	\end{equation}
	\begin{equation} \label{eqn:L-constraint-quantum-2}
		\bra{\phi} \Hat{\widetilde{L}}_{p , k} \ket{\psi} = 0 .
	\end{equation}
\end{subequations}
$\Hat{L}_{p , k}$ has no ambiguities since it's composed of only annihilation operators. As for $\widetilde{L}_{p , k}$, we have an ambiguity for $p = k \neq 0$, so we pick normal ordering with the annihilation operator to the right
\begin{equation} \label{eqn:L-normal-ordering}
	\Hat{\widetilde{L}}_{p , p} = (\Hat{\alpha}^{\mu} _{p})^{\dagger} \Hat{\alpha}^{\nu} _{p} \eta _{\mu \nu} .
\end{equation}
This ambiguity manifests in the imposition of the constraint as
\begin{equation} \label{eqn:shifted-L-constraint}
	\bra{\phi} \left( \sum _{n \neq 0} \left( \left( 1 + \frac{n}{E_{n}} \right) ^2 \Hat{\widetilde{L}}_{n , n} \right) - a \right) \ket{\psi} = 0 ,
\end{equation}
for some constant $a$. Classically, we had eq (\ref{eqn:string-rest-mass-classical-1}) for the string rest mass, from which we can now see that the string mass spectrum will get shifted by this constant
\begin{multline} \label{eqn:shifted-mass-spectrum}
	\Hat{M}^2 = \frac{8}{\alpha '} \Bigg( \frac{2 \mu ^2}{\alpha '} \Hat{x}^{\mu} \Hat{x}_{\mu} - \\
	- \left( \sum _{n \neq 0} \left( \left( 1 + \frac{n}{E_{n}} \right) ^2 E_{n} (\Hat{a}^{\mu} _{n})^{\dagger} \Hat{a}^{\nu} _{n} \eta _{\mu \nu} \right) - a \right) \Bigg) .
\end{multline}
This constant can be calculated by imposing normal-ordering on the summed constraint since $n \in (- d , d)$ is finite, and the result is
\begin{multline}
	a = - D \sum _{n \neq 0} \left( 1 + \frac{n}{E_{n}} \right) ^2 E_{n} = \\
	= - D \sum _{n > 0} \left( 1 + \frac{n^2}{E_{n} ^2} \right) E_{n} ,
\end{multline}
which is easily computable given a value for $\mu$. Below we provide a table with the value of $a$ for different spacetime dimensionalities $D$ and mass parameters $\mu$.
\begin{table}[h] \label{tab:normal-ordering-constant}
	\begin{tabular}{|l|l|l|l|l|l|}
		\hline
		        & $\mu = 1$ & $\mu = 2$ & $\mu = 3$ & $\mu = 4$ & $\mu = 5$ \\ \hline
		$D = 2$ & $-4.24$   & $-5.37$   & $-6.96$   & $-8.73$   & $-10.59$  \\ \hline
		$D = 3$ & $-18.44$  & $-20.78$  & $-24.58$  & $-29.20$  & $-34.27$  \\ \hline
		$D = 4$ & $-48.62$  & $-52.11$  & $-58.23$  & $-66.13$  & $-75.19$  \\ \hline
		$D = 5$ & $-100.79$ & $-105.39$ & $-113.79$ & $-125.09$ & $-138.49$ \\ \hline
		$D = 6$ & $-180.96$ & $-186.63$ & $-197.25$ & $-211.95$ & $-229.84$ \\ \hline
	\end{tabular}
	\caption{values of $a$ for different spacetime dimensionalities and mass parameters. Visibly, the dependence in $D$ is much stronger than the dependence in $\mu$.}
\end{table}

The commutation relations between the constraints are inherited by the Poisson brackets
\begin{subequations}
	\begin{align}
		\left[ \Hat{L}_{n , m} , \Hat{L}_{k , p} \right]                         & = 0 \label{eqn:L-commutator-0}                                                                                                   \\
		\left[ \Hat{L}_{n , m} , \Hat{\widetilde{L}}_{k , p} \right]             & = \Hat{L}_{m , k} \delta _{|n| , |p|} + \Hat{L}_{n , k} \delta _{|m| , |p|} \label{eqn:L-commutator-1}                           \\
		\left[ \Hat{\widetilde{L}}_{n , m} , \Hat{\widetilde{L}}_{k , p} \right] & = \Hat{\widetilde{L}}_{k , m} \delta _{|n| , |p|} - \Hat{\widetilde{L}}_{n , p} \delta _{|m| , |k|} \label{eqn:L-commutator-2} .
	\end{align}
\end{subequations}
The commutator (\ref{eqn:L-commutator-1}) ends up with only annihilation operators in the R.H.S, so it has no ordering ambiguities, meanwhile the commutator (\ref{eqn:L-commutator-2}) have ordering ambiguities for $|m| = |k|$ and/or $|n| = |p|$, so we add the anomalous terms
\begin{multline} \label{eqn:L-commutator-anomalous-initial}
	\left[ \Hat{\widetilde{L}}_{n , m} , \Hat{\widetilde{L}}_{k , p} \right] = \Hat{\widetilde{L}}_{k , m} \delta _{|n| , |p|} - \Hat{\widetilde{L}}_{n , p} \delta _{|m| , |k|} + \\
	+ C_{n} \delta _{|n| , |p|} + D_{k} \delta _{|m| , |k|} .
\end{multline}
Clearly, $C_{0} = D_{0} = 0$, since when all indices are $0$ we have only the position and momentum operators, and thus the commutator should vanish identically. Also, because of the symmetry in the indices we have that $C_{- n} = C_{n}$ and $D_{- k} = D_{k}$. Using the Jacobi identity we get the difference equation
\begin{equation} \label{eqn:F-difference-eqn}
	F_{n + 1} + F_{n} + F_{1} = 0
\end{equation}
for the sum of both sequences $F_{n} = C_{n} + D_{n}$. This difference equation has solution
\begin{equation} \label{eqn:F-sequence}
	F_{n} = \frac{F_{1}}{2} \left( \cos \left( \frac{n \pi}{2} \right) - 1 \right) ,
\end{equation}
and since both $C_{n}$ and $D_{n}$ need to satisfy the initial value of $0$, each of then is half of $F_{n}$,
\begin{equation} \label{eqn:C-D-sequences}
	C_{n} = D_{n} = \frac{F_{n}}{2} = \frac{F_{1}}{4} \left( \cos \left( \frac{n \pi}{2} \right) - 1 \right) .
\end{equation}
Next, by calculating the vacuum expectation value (VEV) of the commutator \eqref{eqn:L-commutator-2} with $|m| \neq |k|$ and $n = p = 1$, we find that $F_{1} = 0$, thus both sequences $C_{n} = D_{n} = 0$ vanish and we conclude that the quantum version of our algebra \emph{has no anomalies}. This is a relevant result, since it could mean our model doesn't need a specific number of dimensions to work, which will be explored again later in section \ref{sec:emergent-dimensions}.

A quick calculation reveals that the WS Hamiltonian and momentum operators are given by
\begin{subequations}
	\begin{align}
		\Hat{H} & = - \frac{1}{2} \sum _{n} \Hat{\widetilde{L}}_{n , n} \label{eqn:WS-Hamiltonian}              \\
		\Hat{P} & = - \frac{1}{2} \sum _{n} \frac{n}{E_{n}} \Hat{\widetilde{L}}_{n , n} \label{eqn:WS-momentum}
	\end{align}
\end{subequations}
since those generate $\tau$ and $\sigma$ translations respectively
\begin{subequations}
	\begin{align}
		\left[ \Hat{H} , \Hat{X}^{\mu} \right] & = - i \del _{\tau} \Hat{X}^{\mu} \label{eqn:tau-generator}     \\
		\left[ \Hat{P} , \Hat{X}^{\mu} \right] & = i \del _{\sigma} \Hat{X}^{\mu} . \label{eqn:sigma-generator}
	\end{align}
\end{subequations}
Clearly, $\Hat{P}$ does not necessarily annihilate physical states (see \eqref{eqn:WS-momentum}), thus those are not required to be invariant under rigid rotations of the string. \emph{Physical states for the KG string have non-conserved WS momentum.} Also, we see that there is no level-matching between left and right-moving modes since those are not distinct in our model.

Let us denote the ground state with momentum $p^{\mu}$ by $\ket{0 ; p}$. Then, the mass-shell condition \eqref{eqn:shifted-mass-spectrum} implies that
\begin{equation} \label{eqn:ground-state-mass-shell}
	\frac{\alpha '}{8} p^2 = \frac{2 \mu ^2}{\alpha '} \left| \frac{\partial \psi (p)}{\partial p} \right| ^2 + a .
\end{equation}
Next we look at the $n$-th excited vector state $\zeta _{\mu} (\Hat{a}^{\mu} _{n})^{\dagger}$ with $\zeta _{\mu} = \zeta _{\mu} (p)$ being the polarization covector. The reason we consider closed string vector states unlike normal String Theory is that in our model, the closed string solution doesn't have distinct left and right sectors, thus permitting spin-1 states. We'll discuss these closed string single excitation states in section \ref{sec:discussion}. The mass-shell now reads
\begin{multline} \label{eqn:nth-state-mass-shell}
	\frac{\alpha '}{8} M^2 = \Bigg( \frac{2 \mu ^2}{\alpha '} \left| \frac{\partial \psi (p)}{\partial p} \right| ^2 - \\
	- \left( 1 + \frac{n^2}{n^2 + \mu ^2} \right) \sqrt{n^2 + \mu ^2} + a \Bigg) ,
\end{multline}
while the auxiliary $\bra{0 ; p} \Hat{\widetilde{L}}_{n , 0} \zeta _{\mu '} (\Hat{a}^{\mu '} _{d})^{\dagger} \ket{0 ; p} = 0$ constraint implies that
\begin{equation} \label{eqn:zeta-p-psi}
	\zeta _{\mu} p^{\mu} = - \frac{4 \mu}{\alpha '} \zeta _{\mu} \frac{\partial \psi (p)}{\partial p_{\mu}} .
\end{equation}
If the momentum wave-function is taken to be
\begin{equation} \label{eqn:p-wave-function}
	\psi (p) = - \frac{\alpha '}{8 \sqrt{2} \mu} p^{\mu} p_{\mu} ,
\end{equation}
eq \eqref{eqn:zeta-p-psi} reduces to $\zeta _{\mu} p^{\mu} = 0$. Putting this back in the mass-shell yields
\begin{equation} \label{eqn:ground-state-mass-shell-simplified}
	\frac{\alpha '}{16} p^2 = a = - D \sum _{k > 0} \left( 1 + \frac{k^2}{E_{k} ^2} \right) E_{k}
\end{equation}
for the ground state and
\begin{multline} \label{eqn:nth-mass-shell-simplified}
	\frac{\alpha '}{16} p^2 = a - \left( 1 + \frac{n^2}{n^2 + \mu ^2} \right) \sqrt{n^2 + \mu ^2} = \\
	= - D \sum _{k > 0} \left( 1 + \frac{k^2}{E_{k} ^2} \right) E_{k} - \left( 1 + \frac{n^2}{E_{n} ^{2}} \right) E_{n}
\end{multline}
for the $n$-th state. The norm of these states is given by $\zeta _{\mu} \zeta ^{\mu}$, and since by eq \eqref{eqn:nth-mass-shell-simplified} we see that $p^2 < 0$, thus $\zeta ^2 > 0$ always holds and we have no ghosts in the closed string spin-1 spectrum. This analysis, however, is for vector states from the closed string. The actual particle we want from the closed string is a symmetric, traceless spin-2 particle, so let's now consider the state $\zeta _{\mu \nu} (\Hat{a}^{\mu} _{n})^{\dagger} (\Hat{a}^{\nu} _{n})^{\dagger} \ket{0 ; p}$ where $\zeta _{\mu \nu}$ is the polarization tensor. The mass-shell reads
\begin{equation} \label{eqn:nth-spin-2-mass-shell}
	\frac{\alpha '}{16} p^2 = - D \sum _{k > 0} \left( 1 + \frac{k^2}{E_{k} ^2} \right) E_{k} - 4 \left( 1 + \frac{n^2}{E_{n} ^{2}} \right) E_{n} .
\end{equation}
The auxiliary $\bra{0 ; p} \xi _{\alpha} \Hat{a}^{\alpha} _{n} \Hat{\widetilde{L}}_{n , 0} (\Hat{a}^{\mu} _{n})^{\dagger} (\Hat{a}^{\nu} _{n})^{\dagger} \zeta _{\mu \nu} \ket{0 ; p} = 0$ condition where $\xi _{\alpha}$ is a arbitrary polarization for the auxiliary vector excitation implies that
\begin{equation} \label{eqn:spin-2-zeta-p=0}
	(\zeta _{\mu \nu} + \zeta _{\nu \mu}) p^{\mu} = 0 .
\end{equation}
We can decompose the polarization into a symmetric traceless part $G_{\mu \nu}$, an anti-symmetric part $B_{\mu \nu}$ and a trace part $\Phi \eta _{\mu \nu}$ given by
\begin{subequations}
	\begin{align}
		G_{\mu \nu} & = \frac{1}{2} (\zeta _{\mu \nu} + \zeta _{\nu \mu}) - \frac{1}{D} \eta ^{\alpha \beta} \zeta _{\alpha \beta} \eta _{\mu \nu} \label{eqn:G-zeta} \\
		B_{\mu \nu} & = \frac{1}{2} (\zeta _{\mu \nu} - \zeta _{\nu \mu}) \label{eqn:B-zeta}                                                                          \\
		\Phi        & = \frac{1}{D} \eta ^{\alpha \beta} \zeta _{\alpha \beta} \label{eqn:Phi-zeta} ,
	\end{align}
\end{subequations}
$\zeta _{\mu \nu} = G_{\mu \nu} + B_{\mu \nu} + \Phi \eta _{\mu \nu}$, giving rise to 3 well-known types of background fields: the spacetime metric $G_{\mu \nu}$, Kalb-Ramond 2-form $B_{\mu \nu}$ and dilaton $\Phi$, all of those having squared mass given by eq \eqref{eqn:nth-spin-2-mass-shell}. We thus see that the symmetric part $S_{\mu \nu} = G_{\mu \nu} + \Phi \eta _{\mu \nu}$ of $\zeta _{\mu \nu}$ is orthogonal to the momentum of the string, $S_{\mu \nu} p^{\mu} = 0$, and since $p^2 < 0$, we can choose a reference frame where $p^{\mu} = (p , 0 , ... , 0)$ only has a time component, leaving us with $S_{0 \nu} = 0$, implying that $\zeta _{0 \nu} = - \zeta _{\nu 0}$. This condition guarantees that the timelike components of $\zeta _{\mu \nu}$ have a positive contribution to its square $\zeta ^{\mu \nu} \zeta _{\mu \nu} = \zeta ^{i j} \zeta _{i j} + b , \ b = \zeta ^{0 i} \zeta _{0 i} > 0$, and the remaining part is spacelike thus must also have positive square, so we get that $\zeta ^{\mu \nu} \zeta _{\mu \nu} > 0$, so we are free of ghosts in the spin-2 spectrum.

It seems we're in trouble since eq \eqref{eqn:nth-spin-2-mass-shell} tells us we have no massless 2 excitation states, but in fact, this is not a problem and is actually expected since quantum geometric effects are expected to violate Lorentz invariance near the Planck scale \cite{Bodendorfer-LQG}, making it so massless particles effectively travel slightly slower than $c$, with corrections on the order of $\sim E_{pl} ^{- 4}$.

The fact that $G_{\mu \nu}$ is symmetric, traceless and satisfies $G_{\mu \nu} p^{\mu} = 0$ leaves it with
\begin{equation} \label{eqn:G-polarizations}
	\frac{1}{2} D (D - 1) - 1
\end{equation}
possible polarizations, which shows that indeed those polarizations fit nicely in a representation of $\mathrm{SO} (D - 1 , 1)$, \ie the traceless symmetric tensor representation, thus all our excited spin-2 states are consistent with Lorentz symmetry at the macroscopic scale even tho near the Planck scale they deviate a bit by having an effective mass.

Although our method is quite different from the one in \cite{Thiemann2004The-LQG-String:}, we got similar results: a quantum Virasoro-like algebra with no anomalies, unconstrained background dimensionality $D$ and no negative-norm states in physical processes.

\section{Emergent Dimensions and LQG} \label{sec:emergent-dimensions}

There is another alternative to incorporate LQG quantum-geometric effects into the string action and get something conceptually similar to what we had in section \ref{sec:inverse-metric-Polyakov}: Afshordi's \textquotedblleft emergent dimension" \cite{Afshordi2014Emergent} model, with the action being
\begin{multline} \label{eqn:emergent-dimensions-action}
	S_{ED} = - \frac{1}{4 \pi \alpha '} \int \diff ^2 x \sqrt{- g} \big( g^{a b} \del _{a} X^{\mu} \del _{b} X^{\nu} \eta _{\mu \nu} + \\
	+ \mu _{0} ^2 \e ^{- m_{\mu} / T} X^{\mu} X^{\nu} \eta _{\mu \nu} \big)
\end{multline}
where $T$ is the \emph{temperature} of target space and the parameters $m_{\mu}$ satisfy an ordering relation $m_{0} \gg m_{1} \gg ... \gg m_{d}$. This is Afshordi's model \cite{Afshordi2014Emergent}. Where our approach diverges is that we realize the parameters $m_{\mu}$ are constrained by the modified dispersion relations of each field $X^{\mu}$, since as \cite{Bodendorfer-LQG} points out, LQG quantum geometry breaks Lorentz invariance near the Planck scale through modifying the dispersion relation by constraining the free parameters of the model as
\begin{equation} \label{eqn:LQG-modified-dispersion-relation}
	E^2 = m^2 + p^2 + \sum _{n \geq 3} c_{n} \frac{p^n}{E_{pl} ^{n - 2}} ,
\end{equation}
so we note that the mass parameters of the fields $X^{\mu}$ will be constrained as
\begin{equation} \label{eqn:X-masses-constraint}
	E_{\alpha} ^2 = p_{\alpha} ^2 + \sum _{n = 3} ^{D - 1} \frac{m_{n}}{T} \frac{p_{\alpha} ^{n}}{E_{pl} ^{n - 2}} ,
\end{equation}
where $E_{\alpha} ^2$ does not mean $E_{\alpha} E^{\alpha}$, rather it means $(E_{3})^2 , (E_{4})^2 , ... , (E_{D - 1})^2$, \ie, the squared WS energy (and momentum) of each field $X^{\mu}$.

Afshordi's work notes that the action \eqref{eqn:emergent-dimensions-action} only has background Poincaré invariance and WS Weyl symmetry as approximate symmetries for the fields with $m_{\mu} \gg T$, thus the \textquotedblleft emergent spacetime and evolving dimensions" by interpreting the fields with effective mass as being inaccessible due to not obeying Poincaré symmetry. We go one step further and note that since eq \eqref{eqn:X-masses-constraint} leaves $m_{0} , m_{1}$ and $m_{2}$ unconstrained, the related dimensions might be regarded as being \emph{fundamental} and thus the true quantum gravitational field should be $(2 + 1)$-dimensional, in agreement with \cite{t-Hooft1993Dimensional}. $(2 + 1)$-dimensional gravity has already been successfully quantized before (\cite{Bergshoeff2013ZweiDreibein},\cite{Blais2013A-new-quantization},\cite{Carlip2003Quantum}). We thus postulate that $m_{3} \sim T_{pl}$, which together with the constraints \eqref{eqn:X-masses-constraint} should give us a way to calculate the other mass parameters given the WS energy and momentum of the fields, making it possible to predict in which scales the other dimensions should start to open up.

By analysing the equations of motion for the metric $g_{c d}$ from the action \eqref{eqn:emergent-dimensions-action} we get
\begin{equation} \label{eqn:emergent-dimensions-g}
	g_{c d} = 2 f \del _{c} X^{\mu} \del _{d} X^{\nu} \eta _{\mu \nu} ,
\end{equation}
with $f$ given by
\begin{equation} \label{eqn:emergent-dimensions-f}
	\frac{1}{f} = g^{a b} \del _{a} X^{\mu} \del _{b} X^{\nu} \eta _{\mu \nu} + \mu _{0} ^{2} \e ^{- m_{\mu} / T} X^{\mu} X^{\nu} \eta _{\mu \nu} ,
\end{equation}
from which we see that $\e ^{- m_{\mu} / T} X^{\mu} X^{\nu} \eta _{\mu \nu} = 0$ is actually a constraint and thus the base mass $\mu _{0} ^2$ is a Lagrange multiplier. By dimensional analysis, we see that in SI units $\mu _{0} ^2$ has units of inverse area, so it could be that $\mu _{0} ^2 \sim A_{j} ^{-1}$, where $A_{j}$ is the eigenvalue of the area operator with spin $j$, which depends on where the spin network punctures the WS, thus once more connecting Afshordi's work to LQG quantum geometry.

One could also consider a more general form of action \eqref{eqn:emergent-dimensions-action} by swapping the flat Minkowski metric $\eta _{\mu \nu}$ by a more general curved metric $G_{\mu \nu}$ and considering off-diagonal mass parameters $m_{\mu \nu}$:
\begin{multline} \label{eqn:emergent-dimensions-curved}
	S' _{E D} = - \frac{1}{4 \pi \alpha '} \int \diff ^2 x \sqrt{- g} \big( g^{a b} \del _{a} X^{\mu} \del _{b} X^{\nu} G_{\mu \nu} + \\
	+ \mu _{0} ^2 \e ^{- m_{\mu \nu} / T} X^{\mu} X^{\nu} G_{\mu \nu} \big) ,
\end{multline}
where the mass parameters will be constrained as in
\begin{equation} \label{eqn:X-masses-constraint-2}
	E_{\alpha} ^2 = p_{\alpha} ^2 + \sum _{n = 3} ^{D - 1} \frac{m_{n \alpha}}{T} \frac{p_{\alpha} ^n}{E_{pl} ^n} ,
\end{equation}
since those are the ones that show up in the $X^{\mu}$ equations of motion
\begin{multline}
	g^{a b} \nabla _{a} \nabla _{b} X^{\mu} G_{\mu \lambda} - \mu _{0} ^2 \e ^{- m_{\mu \lambda} / T} X^{\mu} G_{\mu \lambda} = \\
	= \frac{1}{2} g^{a b} \del_{a} X^{\mu} \del _{b} X^{\nu} \del _{\lambda} G_{\mu \nu} .
\end{multline}

In section \ref{sec:inverse-metric-Polyakov} we arrived at eq \eqref{eqn:coupling-X-phi} and claimed the coupling between the WS conformal factor $\phi (x)$ and the embedding fields $X^{\mu}$ to be akin to the dilaton field $\Phi (X)$. We now explore this claim: since we could be dealing with emergent dimensions with an action \eqref{eqn:emergent-dimensions-action} that only has Weyl symmetry as an approximate symmetry, we believe that the conformal factor $\phi (x)$ can be viewed as a \emph{worldsheet dilaton}, which sets the scale of the string dictating which dimensions it can interact with, thus the coupling \textquotedblleft constant" of the masses may be related to the conformal factor
\begin{equation} \label{eqn:mu0-phi-relation}
	\mu _{0} ^2 \sim \e ^{- \phi (x)} ,
\end{equation}
where the minus sign is due to the fact that the dimensions open up at larger and larger scales, so we require a larger and larger string to interact with each new dimension.

\section{Holography From Classical Strings}\label{sec:holographic-strings}

Until now we have studied a phenomenological modification of the string action. However, ultimately all phenomenological ideas must have a grounding in some exact, microscopic description of the system. Further, we have neglected one very important aspect of this whole endeavor, which is the description of the background geometry in the connection formalism and how to couple that geometry to the geometry of the string worldsheet. In this section, we present a possible action for the bosonic string in which the effect of the bulk geometry on the worldsheet is encoded in the string action in terms of a connection which lives on the worldsheet but takes values in the Lie algebra of the gauge group which is being used to describe the bulk geometry.  We will explore the natural consequences which follow from considering the symmetries of such an action.

First let us recall the form of the Polyakov action for the bosonic string \eqref{eqn:polyakov-action}:
\begin{equation*}
	S_{P} = -\frac{T}{2} \int \diff \tau \wedge \diff \sigma \sqrt{-g} g^{ab} \partial_a X^\mu \partial_b X^\nu \eta_{\mu\nu}.
\end{equation*}
If one was to ask a beginning graduate student as to how one can incorporate a gauge field in the above expression, the natural answer would be that we can promote the $X^\mu$ to vectors which live in the representation of some gauge group $\mc G$ and promote the ordinary derivative to a gauge covariant derivative \wrt a connection valued in $\mc G$. The resulting action would now have the form:
\begin{equation}\label{eqn:polyakov-gauged}
	S_{P} = -\frac{T}{2} \int \diff \tau \wedge \diff \sigma \sqrt{-g} g^{ab} D_a X^\mu D_b X^\nu \eta_{\mu\nu},
\end{equation}
where the covariant derivative $D_a$ is of the form:
\begin{equation}\label{eqn:covariant-deriv-v1}
	D_a X^\mu(\tau , \sigma) = \partial_a X^\mu + \mu \mc{A}_{a}^\mu{}_\nu X^\nu,
\end{equation}
where $\mc{A}_{a}^\mu{}_\nu$ is the gauge field and $k$ is the gauge coupling. A more experienced researcher would object saying that we know that the $X^\mu$ are spacetime co-ordinates and not vectors transforming under the gauge group $\mc G$. But we will see that this is not a problem. In fact, in order to be able to couple the string to an \emph{arbitrarily} curved geometry we will find that we \emph{require} the $X^\mu$ to transform as vectors under the action of some \emph{local} gauge field.

Our goal is to write down the action of a string propagating in a background geometry. One might argue that the expression \eqref{eqn:polyakov-curved-bulk}, where we replace the bulk Minkowski tensor $\eta_{\mu\nu}$ with an arbitrary tensor $G_{\mu\nu}$ which tells us what the components of the bulk metric are at that point on the worldsheet already describes the string in an arbitrary background geometry. However, we view this approach as unsatisfactory, not the least because it does not allow us to make contact of the string action with a background geometry specified in terms of connection variables instead of metric ones. An astute reader might correctly object saying that we can simply replace $G_{\mu\nu}$ in \eqref{eqn:polyakov-curved-bulk} with $E_\mu^I E_\nu^J \eta_{IJ}$ (here $\eta_{IJ}$ is now the Minkowski metric in locally flat coordinates). This, however, leaves open the question of what role the bulk connection should play in this picture.

Further, we can ask what is the physical effect of having a non-trivial connection in the bulk in terms of the local geometry. The bulk connection tells us how the vielbein transforms from point to point and thus effectively measures how much the bulk metric changes from point to point. So, for instance, the connection which describes the geometry around a Kerr black hole also encodes how the local tetrad frame rotates as one moves around the axisymmetric axis of the black hole. All of these considerations lead us to the premise that \emph{if one wishes to describes the bulk geometry in terms of connection variables, then it is only natural to describe the worldsheet geometry in terms of connection variables} and to use the tools that come along with having a gauge connection to understand how the worldsheet geometry can couple to the bulk geometry.

Therefore, now we will require that the bulk connection $\mc{A}_{\mu} ^{I J}$ and the bulk vielbein $E^{\mu} _{I}$ can be pulled back onto a connection $\mc{A}_{a}^{I J}$ and a dyad $e^a_I$, respectively, living on the string worldsheet $\Sigma$. $X^\mu(\tau , \sigma)$ are the space-time coordinates of a point $(\tau , \sigma)$ on the worldsheet. Now if one adopts the ``sigma model'' point of view then the $X$'s can be viewed as $d+1$ scalar fields living on a $1+1$ worldsheet. But we know that these fields are not a collection of $d+1$ independent scalars. They are components of a $d+1$ dim vector since they transform with the fundamental representation of $\mathrm{SO}(d,1)$ under global Lorentz transformations acting on the ambient space-time. However, the symmetries of a flat background include not only Lorentz transformations bulk also translations. The string action is required to be invariant under both of these:
\begin{subequations}\label{eqn:poincare-symmetries}
	\begin{align}
		X^\mu & \rarrow X^\mu + \delta X^\mu; \quad \delta X^\mu \in \mbb{R}^{3,1}            \\
		X^\mu & \rarrow \Lambda^\mu{}_\nu X^\mu; \quad \Lambda^\mu{}_\nu \in \mathrm{SO}(3,1)
	\end{align}
\end{subequations}
where $\delta X^\mu$ is some arbitrary vector in Minkowski space. The semidirect product of $\mbb{R}^{3,1}$ with $\mathrm{SO}(3,1)$ is $ISO^{3,1}$, the Poincaré group in $3+1$ dimensions. Of course, these are good symmetries only if our background is flat. An arbitrarily curved background can have one or more Killing vectors depending on how much symmetry the geometry has, but \emph{in general} even that need not be the case. The best one can hope for is some notion of local flatness is some finite subregion of the background manifold. Moreoever the entire purpose of this work to try and understand how to write down a string action in an \emph{arbitrarily curved} background, assuming that such a thing is even possible.

Even in an arbitrarily curved background, however, symmetry still plays a very important role. Even if we have discarded \emph{global} symmetries, that still leaves us with \emph{local} symmetries in the form of diffeomorphisms under which the action must remain invariant. If, in addition, we have fields such as a gauge connection and an associated vielbein living in the background, then we also have to take local gauge transformations of these objects into account.

However, diffeomorphisms and gauge transformations are not distinct concepts. A diffeomorphism is a differentiable map from one manifold to another (in the case of general relativity both manifolds are generally taken to be the same), which is a differentiable (and invertible, with inverse also differentiable) map. In more pedestrian language, a diffeomorphism from $\mc{M}$ to itself is an (invertible) change of coordinates on a manifold $\mc{M}$. A gauge transformation is a redefinition of the local fields which live on the underlying manifold. In fact, at least for certain geometries, these two can be seen to be equivalent to each other, with diffeomorphisms representing ``passive'' coordinate transformations and gauge representing ``active'' coordinate transformations. From the viewpoint of the formalism of Principal Fiber Bundles, where the gauge symmetry arises from a structure that locally looks like a Cartesian product of the base manifold $\mc{M}$ and the gauge group $G$, $\mc{M} \times G$, gauge transformation arise from local reparameterizations of the group component at each point of the base manifold, so it is a diffeomorphism in the \textquotedblleft enlarged" space $\mc{M} \times G$.

We will assume henceforth that the Poincaré group is the gauge group which must be promoted from being a \emph{global} symmetry to being a \emph{local} symmetry of the background spacetime.

\subsection{Gauging Lorentz Transformations}\label{sec:gauging-lt}

Instead of viewing the fields $X^{\mu}$ as elements of a Minkowski space, we think of them as elements of the tangent space $T_p (M)$ at a given point $p$ of the background manifold $M$.

This viewpoint on the embedding fields, as elements of the tangent space $T_p(M)$, might appear to be at odds with their prescribed usage as co-ordinate functions on $M$. However, some reflection shows that this is not the case. For instance, one can take the embedding manifold to be a two-dimensional sphere $S^2$. Then the coordinates of a point on the string worldsheet $\Sigma$ can be expressed in terms of the polar and azimuthal angles $\theta$ and $\phi$. The action for this system would be:
\begin{equation}\label{eqn:s2-string}
	S_{S^2} = -\frac{T}{2} \int \diff^2 x \sqrt{-g} g^{ab} \partial_a \theta^\mu \partial_b \theta^\nu G_{\mu\nu},
\end{equation}
where $\theta^\mu(\sigma,\tau) \in \{\theta, \phi\}$ are the embedding fields, and $G_{\mu\nu}$ is \emph{a} metric living on $S^2$.

There is a difficulty in the above argument. The bulk geometry is $d+1$ dimensional, therefore there are $d+1$ embedding fields. We have taken these fields $X^\mu$ and cast them as components of a $\mf{so}(d,1)$ vector. Let us pick an arbitrary point $P_0$ on the bulk manifold and call it the ``origin''. At $P_0$ the co-ordinates are all zero $\vec X_0 = (0,0,\ldots,0)$. If we start acting with infinitesimal $\mf{iso}(d,1)$ transformations on $\vec X_0$, the resulting surface will correspond to a ``constant energy'' hyperboloid in the bulk. This is because $SO(d,1)$ contains only boosts and rotations. We also need translations to generate the entirety of the bulk starting from a single base point.

\subsection{Gauging Translations}\label{sec:translation_gauge}

This is where we run into a problem. There is no way to express infinitesimal translations in $d+1$ dimensions as the action of a $d+1$ dimensional matrix-valued group on a point living in $\mbb{R}^{d+1}$. In order to incorporate translations we have to resort to a trick and introduce an extra ``dummy'' dimension for the background manifold. Now we have $d+2$ embedding fields which we can write as:
\begin{equation}\label{eqn:enlarged-Xs}
	X^A := \{X^0, \vec X, X^r\} = \{X^\mu, X^r\},
\end{equation}
where $X^\mu = \{X^0, \vec X\}$ are the fields we had originally and $X^r$ is our new ``dummy'' co-ordinate. As we will see this dimension will turn out to have a very important physical interpretation in the context of Einstein-Cartan geometry.

If one wants to treat the $X$'s as elements of a vector which transforms under both translations and rotations as in \eqref{eqn:poincare-symmetries}, we must extend the number of spacetime dimensions by one, introducing an ``auxiliary'' dimension whose value is initially set to be $X^r = 1$. Thus we have:
\begin{equation}
	\begin{pmatrix}
		X^\mu \\
		1
	\end{pmatrix} \rarrow \begin{pmatrix}
		X'^\mu \\
		1
	\end{pmatrix} = \begin{pmatrix}
		\Lambda^\mu{}_\mu & \delta X^\mu \\
		0                 & 1
	\end{pmatrix} \begin{pmatrix}
		X^\mu \\
		1
	\end{pmatrix}
\end{equation}
Now, one might be inclined to treat the new co-ordinate $X^r$ as purely an auxiliary variable which will not affect the physics. However, if we stare at the expression $X^r = 1$ for a moment, we will realise that this expression is the same form as the Dirichlet boundary condition for the end-points of a string. And, from what we know about strings with Dirichlet boundary conditions is that the constraint surface on which the end-points live will end up behaving like a dynamical object or ``D-brane'' \cite{Dai1989New-Connections,Polchinski1995Dirichlet-Branes,Horava1989Background}. In our case we now have a new enlarged bulk $\mb{M}$ with $d+1$ space dimensions - treating the auxiliary variable as a co-ordinate which can vary - and one time dimensions, or $d+2$ dimensions in total. If we fix $X^r = 1$, that leaves us with a $d+1$ submanifold $\mb{\Gamma} \in \mb{M}$ on which the string end-point is free to move.

However, in our case, the condition $X^r = 1$ is much stronger than the Dirichlet boundary condition, because it applies not only to the end-points of the string, but to every point of the worldsheet. In other words, the \emph{entire} string worldsheet is confined to move only along the $d+1$ dimensional submanifold $\mb{\Gamma}$. Now, one can ask what should the form of the action for a string moving in this enlarged bulk look like? Well, it will be of the same form \eqref{eqn:polyakov-gauged}, only now the worldsheet connection is an $\mf{iso}(d,1)$ valued one-form, with the embedding fields living in the enlarged $d+2$ dimensional set.

\subsection{The Einstein-Cartan String}\label{sec:einstein-cartan-string}

It is useful to see the explicit form of a $\mf{iso}(3,1)$ gauge connection. The elements of the Lie algeba of $\mf{iso}(3,1)$ have the following form \cite{Wise2010MacDowell-Mansouri}
\begin{align}\label{eqn:poincare-lie-algeba}
	T^A{}_B & := \begin{pmatrix}
		             0          & b^1        & b^2        & b^3        & p^0/l \\
		             b^1        & 0          & j^3        & -j^2       & p^1/l \\
		             b^2        & -j^3       & 0          & j^1        & p^2/l \\
		             b^3        & j^2        & -j^1       & 0          & p^3/l \\
		             \ep p^0/l & -\ep p^1/l & -\ep p^2/l & -\ep p^3/l & 0
	             \end{pmatrix} \nonumber \\
	        & = j^i J_i + b^i B_i + \frac{1}{l}p^a P_a,
\end{align}
where $J_i, B_i$ are the generators of rotations and boosts respectively. $P_a := (P_0, P_i)$ are the generators of translations. The parameter $\ep$ takes values in $\{-1,0,1\}$. Strictly speaking the above expression corresponds to an $\mf{iso}(3,1)$ Lie algebra generator only when $\ep = 0$. $l$ is a length scale which must be introduced so that the components of the ``vector'' $(p^0, p^i)$ can be viewed as components of a translation vector along the $3+1$ dimensional geometry $\mb{\Gamma}$. Similarly the boosts and rotations describe
where $J_i, B_i$ are the generators of rotations and boosts respectively. $P_a := (P_0, P_i)$ are the generators of translations. The parameter $\ep$ takes values in $\{-1,0,1\}$.

Strictly speaking the above expression corresponds to an $\mf{iso}(3,1)$ Lie algebra generator only when $\ep = 0$. The case when $\ep = \pm 1$ will be motivated and discussed later. $l$ is a length scale which must be introduced so that the components $(p^0, p^i)$ can be viewed as components of a ``translation vector'' living in the tangent space of the $3+1$ dimensional geometry $\mb{\Gamma}$. Similarly the boosts and rotations generators are for objects living in $\mb{\Gamma}$.

Let us write down the gauged Polyakov action for the Poincaré lie algebra valued connection $\mc{A}_a$ whose components are given in \eqref{eqn:poincare-lie-algeba}:
\begin{equation}\label{eqn:poincare-polyakov}
	S'_{P} = -\frac{T}{2} \int \diff^2 x \sqrt{-g} g^{ab} \mc{D}_a X^A \mc{D}_b X^B G_{AB},
\end{equation}
where now the $X^A = \{X^\mu, X^r\}$ (as mentioned previously in \eqref{eqn:enlarged-Xs}) and $G_{AB}$ is the expanded metric on the $d+2$ enlarged bulk manifold $\mb{M}$ and the covariant derivative is of the form defined in \eqref{eqn:covariant-deriv-v1}, but only now acting on $X^A$ instead of on $X^\mu$:
\begin{equation}\label{eqn:covariant-deriv-v2}
	\mc{D}_a X^A(\tau , \sigma) = \partial_a X^A + \mu \mc{A}_{a}^A{}_B X^B.
\end{equation}
Now, we will expand out the action \eqref{eqn:poincare-polyakov} and express it in terms of the $X^\mu$ and $X^r$ separately. It is clear that the resulting expression will be sum of three terms, with the first term being the (gauged) Polyakov action for the string on $\mb{\Gamma}$, terms describing the coupling of the string on $\mb{\Gamma}$ with fluctuations along the extra dimensions $X^r$ and finally a Polyakov like term for the lone extra dimension. We can write this schematically as:
\begin{equation}\label{eqn:poincare-polyakov-decomp}
	S'_P[X^A] = S_P[X^\mu] + S_{int}[X^\mu, X^r] + S_P[X^r] ,
\end{equation}
or more explicitly as
\begin{multline} \label{eqn:poincare-polyakov-decomp-explicit}
	S' _{P} [X^{\mu} , X^{r}] = -\frac{T}{2} \int \diff ^2 x \sqrt{- g} g^{a b} \mc{D}_{a} X^{\mu} \mc{D}_{b} X^{\nu} \eta _{\mu \nu} - \\
	- T \int \diff ^2 x \sqrt{- g} g^{a b} \mc{D}_{a} X^{\mu} \mc{D}_{b} X^{r} G_{\mu r} - \\
	- \frac{T}{2} \int \diff ^2 x \sqrt{- g} g^{a b} \mc{D}_{a} X^{r} \mc{D}_{b} X^{r} G_{r r} ,
\end{multline}
where we can see a natural decomposition of the expanded bulk metric $G_{A B}$ as a $(d+1)$-dimensional tensor $\eta _{\mu \nu}$, a $(d+1)$-dimensional vector $V_{\mu} = G_{\mu r}$ and a scalar $V = G_{r r}$, akin to the ADM decomposition of the metric in standard GR literature.

One can now argue along the same lines as the original arguments which led to the discovery of D-branes \cite{Dai1989New-Connections,Polchinski1995Dirichlet-Branes,Horava1989Background} to say that the extra dimension $X^r$ will become dynamical in the full theory even though it began life only as a classical constraint on the string worldsheet's location. This will imply that while most of the time the string will remain confined to $\mb{\Gamma}$, fluctuations along $X^r$ can cause the string to leave the ``brane'' and venture out into the bulk $\mb{M}$. If this is truly the case then the geometry of the enlarged bulk $\mb{M}$ will be essential for a description of the full string dynamics.

This is where the non-zero value of $\ep$ mentioned in \eqref{eqn:poincare-lie-algeba} come into play. It turns out that for $\ep = +1$ ($\ep = -1$) the connection in \eqref{eqn:poincare-lie-algeba} describes a $\mf{so}(4,1)$ ($\mf{so}(3,2)$) connection respectively. These describe the symmetries of (asymptotically) deSitter and anti-deSitter spacetimes respectively. The length scale $l$ determines the value of the effective cosmological constant in the dS (adS) geometries:
\begin{equation}\label{eqn:cosmological-constant}
	\lambda = \frac{3 \ep}{l^2}.
\end{equation}
One might ask what is the motivation for introducing dS (adS) geometries into the discussion. The main motivation of course is the Maldacena or adS/CFT correspondence. The adS spacetime is usually understood to arise as a consequence of non-perturbative effects which nobody really has a proper understanding of as yet. One might hope that for a correspondence which has passed all possible checks uptil now, there should be some way to realise it in the classical theory also.

What our discussion in this section shows, however, is that such a classical description is possible if we switch from the metric based approach to the connection based one for describing the background geometry and its coupling to the string worldsheet.

\subsection{Gauged Polyakov Equations of Motion}

By varying the action \eqref{eqn:poincare-polyakov}, we get two sets of constraints from the variation of $g_{a b}$ and $A_{a} ^{A B}$, namely the vanishing of the worldsheet energy-momentum and spin-linear-momentum tensors $T_{a b}$ and $\sigma _{a} ^{A B}$
\begin{multline} \label{eqn:gauged-WS-EM-tensor}
	T_{a b} : = \mc{D}_{a} X^{A} \mc{D}_{b} X^{B} G_{A B} - \\
	- g_{a b} g^{c d} \mc{D}_{c} X^{A} \mc{D}_{d} X^{B} G_{A B} = 0
\end{multline}
\begin{equation}
	\sigma ^{A B} _{a} : = \mc{D}_{a} X^{[A} X^{B]} = 0 .
\end{equation}
The former allows us to solve for the worldsheet \textquotedblleft covariant metric" $g_{a b}$ as
\begin{equation} \label{eqn:gauged-WS-metric}
	g_{a b} = 2 f \mc{D}_{a} X^{A} \mc{D}_{b} X^{B} G_{A B}
\end{equation}
where
\begin{equation} \label{eqn:gauged-f-metric}
	\frac{1}{f} = g^{a b} \mc{D}_{a} X^{A} \mc{D}_{b} X^{B} G_{A B} ,
\end{equation}
while the latter should allow one to solve for $\mc{A}_{a}$ in terms of $X^{A}$.

The variation of the action \wrt $X^{A}$ results in the equation of motion of the embbeding fields
\begin{multline} \label{eqn:gauged-X-EoM}
	\mc{D}_{a} [\sqrt{- g} g^{a b} \mc{D}_{b} X^{B} G_{B C}] = \\
	= \sqrt{g} g^{a b} \mc{D}_{a} X^{A} \mc{D}_{b} X^{B} \partial _{C} G_{A B} ,
\end{multline}
which in worldsheet \textquotedblleft covariant conformal gauge" $g^{a b} = \e ^{- \phi} \eta ^{a b}$ and for flat background geometry $G_{A B} = \eta ' _{A B}$ with
\begin{equation} \label{eqn:flat-bulk-metric}
	[\eta ' _{A B}] = \begin{bmatrix} [\eta _{\mu \nu}] & 0 \\ 0 & \epsilon \end{bmatrix}
\end{equation}
simplifies to
\begin{equation}
	\eta ^{a b} \mc{D}_{a} \mc{D}_{b} X^{A} \eta ' _{A B} = 0 .
\end{equation}
From here on, we'll work on the case of $\ep = - 1$ (adS background), for which the above equation can be expanded as
\begin{multline} \label{eqn:expanded-gauged-X-EoM}
	\eta ^{a b} \big( \del _{a} \del _{b} X^{A} + 2 \mu A^{A} _{a B} \del _{b} X^{B} + \\
	+ \mu \del _{a} A^{A} _{b B} X^{B} + \mu ^2 A^{A} _{a B} A^{B} _{b C} X^{C} \big) = 0 .
\end{multline}
To produce the desired bulk geometry, the background connection can be taken as
\begin{equation} \label{eqn:adS-bulk-connection}
	\mc{A}_{\mu} = \begin{bmatrix} 0 & 0 & 0 & 0 & \delta ^{0} _{\mu} \\ 0 & 0 & 0 & 0 & \delta ^{1} _{\mu} \\ 0 & 0 & 0 & 0 & \delta ^{2} _{\mu} \\ 0 & 0 & 0 & 0 & \delta ^{3} _{\mu} \\ - \delta ^{0} _{\mu} & \delta ^{1} _{\mu} & \delta ^{2} _{\mu} & \delta ^{3} _{\mu} & 0 \end{bmatrix} ,
\end{equation}
from which the pulledback connection $\mc{A}_{a} = \del _{a} X^{\mu} \mc{A}_{\mu}$ is thus
\begin{equation} \label{eqn:adS-WS-connection}
	\mc{A}_{a} = \begin{bmatrix} 0 & 0 & 0 & 0 & \del _{a} X^{0} \\ 0 & 0 & 0 & 0 & \del _{a} X^{1} \\ 0 & 0 & 0 & 0 & \del _{a} X^{2} \\ 0 & 0 & 0 & 0 & \del _{a} X^{3} \\ - \del _{a} X^{0} & \del _{a} X^{1} & \del _{a} X^{2} & \del _{a} X^{3} & 0 \end{bmatrix} ,
\end{equation}
whereby imposing conformal gauge on the induced metric $\gamma _{a b} = \del _{a} X^{\mu} \del _{b} X^{\nu} \eta _{\mu \nu} = \e ^{\varphi} \eta _{a b}$ we can separate equations for the Minkowski sector $X^{\mu}$ and for the holographic dimension $X^{r}$:
\begin{multline} \label{eqn:Minkowski-sector-EoM}
	\eta ^{a b} \del _{a} \del _{b} X^{\mu} + \\
	+ \frac{\mu \eta ^{a b}}{1 + \mu X^{r}} \left( 2 \del _{a} X^{r} + k \del _{a} X^{\nu} X_{\nu} \right) \del _{b} X^{\mu} = 0
\end{multline}
\begin{multline} \label{eqn:holographic-dim-EoM}
	\eta ^{a b} \del _{a} \del _{b} X^{r} + 2 \mu ^2 \e ^{\varphi} X^{r} + 4 \mu \e ^{\varphi} - \\
	- \frac{\mu ^2 \eta ^{a b}}{1 + \mu X^{r}} \left( 2 \del _{a} X^{r} + \mu \del _{a} X^{\nu} X_{\nu} \right) \del _{b} X^{\mu} X_{\mu} = 0 .
\end{multline}
Assuming the perturbations on the string are small $|X^{A}| \ll 1$, we get a massive simplification
\begin{equation} \label{eqn:simplified-Minkowski-EoM}
	\eta ^{a b} \del _{a} \del _{b} X^{\mu} = 0
\end{equation}
\begin{equation} \label{eqn:simplified-holographic-EoM}
	\eta ^{a b} \del _{a} \del _{b} X^{r} + 2 \mu ^2 \e ^{\varphi} X^{r} = - 4 \mu \e ^{\varphi} .
\end{equation}
Thus, the Minkowski sector behaves like regular ST for small perturbations, and for the holographic dimension we can solve using a Green's function
\begin{equation} \label{eqn:Xr-GF}
	X^{r} (x) = - 4 \mu \int \diff ^2 x' G (x ; x') \e ^{\varphi (x')} ,
\end{equation}
with the Green's function satisfying
\begin{equation} \label{eqn:Greens-function}
	(\del _{\sigma} ^2 - \del _{\tau} ^2 + 2 \mu ^2 \e ^{\varphi}) G (x ; x') = \delta ^2 (x , x') .
\end{equation}
To find this Green's function, we must fix the conformal factor to be of the form $\e ^{\varphi} = f (\tau) - g (\sigma)$, thus allowing us to separate variables in the homogeneous case of eq \eqref{eqn:Greens-function} into
\begin{equation} \label{eqn:Green-X}
	X^{\prime \prime} - 2 \mu ^2 g (\sigma) X = - \lambda X
\end{equation}
\begin{equation} \label{eqn:Green-T}
	\ddot{T} - 2 \mu ^2 f (\tau) T = - \lambda T .
\end{equation}
These are Schrodinger's equations for potentials given by $\mu f (\tau)$ and $\mu g (\sigma)$, as such if we assume that $f$ and $g$ are harmonic oscillator potentials $f (\tau) = \omega ^2 \tau ^2 / 2$ and $g (\sigma) = k^2 \sigma ^2 / 2$ we can find known solutions for separation constant $\lambda = 2 \mu E = \mu k (2n + 1) = \mu \omega (2n + 1)$, so by the Spectral Theorem our separated spatial and temporal Green's functions are
\begin{equation} \label{eqn:X-function}
	X (\sigma , \sigma ') = \sum _{n \geq 0} - \frac{X_{n} (\sigma) X_{n} (\sigma ')}{\mu k (2 n + 1)}
\end{equation}
\begin{equation} \label{eqn:T-function}
	T (\tau , \tau ') = \sum _{m \geq 0} - \frac{T_{m} (\tau) T_{m} (\tau ')}{\mu \omega (2 m + 1)} ,
\end{equation}
where $X_{n}$ and $T_{m}$ are harmonic oscillator eigenfunctions. With this, the full Green's function $G = T.X$ and the solution for $X^{r}$ are thus
\begin{equation} \label{eqn:G-function}
	G (x ; x') = \sum _{m \geq 0} \sum _{n \geq 0} \frac{T_{m} (\tau) T_{m} (\tau ')}{\mu \omega (2 m + 1)} \frac{X_{n} (\sigma) X_{n} (\sigma ')}{\mu k (2 n + 1)}
\end{equation}
\begin{multline} \label{eqn:Xr-solution}
	X^{r} (x) = - 4 \mu \sum _{m \geq 0} \sum _{n \geq 0} \frac{T_{m} (\tau)}{\mu \omega (2 m + 1)} \frac{X_{n} (\sigma)}{\mu k (2 n + 1)} \times \\
	\times \int \diff ^2 x' T_{m} (\tau ') X_{n} (\sigma ') \left( \frac{\omega ^2 \tau ^{\prime 2}}{2} - \frac{k^2 \sigma ^{\prime 2}}{2} \right) .
\end{multline}



\section{Discussion} \label{sec:discussion}

In this article we have presented a phenomenological modification of the Nambu-Goto action \emph{inspired by the central result of loop quantum gravity, according to which area and other geometric observables are quantized at the Planck scale}. In particular the spectrum of the area operator is discrete and the smallest eigenvalue is non-zero (of the order $l_{pl}^2 \sim 10^{-68}\,\text{m}^2$). On the other hand we also took as a working hypothesis the statement that string theory, or some suitable version of it, provides the correct description of the low energy physics of the world around us. We then searched for the way in which both of these paradigms can be incorporated into a single theoretical model. The form of the action shown in \eqref{eqn:ng-area-corrected} appears, at least superficially, as the simplest possible modification of the Nambu-Goto action which incorporates a minimum eigenvalue for the area. This value is encoded in the modulus of the $\Delta(x)$ field. In the process we introduced a new coupling parameter $ k = l_{pl}/l_s$ which is the ratio of the string scale to the Planck scale.

Assuming that $l_{pl} \ll l_s$, we have $k \ll 1$ and consequently \eqref{eqn:ng-area-corrected} can be expanded as a power series in $k$. Keeping only the term linear in $k$, we obtain an action which is the sum of two copies of the Nambu-Goto action, one for the original metric $h_{ab}$ and the second for the inverse metric $h^{ab}$. We then pointed out the similarity between this expression and previous studies of \emph{bimetric gravity} \cite{Hassan2012Bimetric}. We also want to point that though the possible relationship between string theory and bimetric gravity has been investigated before by \cite{Lust2021Extracting,Yan2021Strings}, these authors have looked at bimetric gravity in the \emph{bulk} geometry rather than a bimetric action for the worldsheet, itself, of the type given in \eqref{eqn:modified-ng-action-v2}.

The bimetric viewpoint for the area-corrected Nambu-Goto action is more than just an interesting curiosity. This perspective allows us to construct the Polyakov action for the bimetric string in \eqref{eqn:polyakov-bimetric}. Of course, one would expect that the (area-corrected) Polaykov and Nambu-Goto formulations should be physically equivalent. Imposing this requirement leads us to the conclusion that the object $\Delta$ in \eqref{eqn:ng-area-corrected} must be worldsheet two-form (antisymmetric tensor) corresponding to the pullback of the \emph{bulk} Kalb-Ramond field. All of this eventually leads us to recognise the relation of our proposed action \eqref{eqn:ng-area-corrected} to the Born-Infeld action \eqref{eqn:BI-action} which arises when one studies the dynamics of D-branes in string theory.

Let us make one crucial observation here. In the expression \eqref{eqn:ng-area-corrected} we originally took $\Delta$ to a constant representing the smallest quantum of area. It is only \emph{later}, by demanding consistency of the area-corrected Nambu-Goto and Polaykov actions that we realise that $\Delta \equiv \Delta_{ab}(x) $ is in fact an antisymmetric worldsheet two-form!

After exploring the equations of motion of the area corrected string, we found that said equations can be recast as a set of Klein-Gordon equations, which for a first analysis we assumed to have constant coefficient $\mu$, however that need not be the case. The solution of these equations and their consequences for the constraints lead us to a modified \textquotedblleft Virasoro-Klein-Gordon" algebra, where the algebra modes have two indices, which in turn resulted in a truncated solution. This also had the effect of not having distinct left and right moving modes, thus there was no level-matching prohibiting spin-1 states for the closed string, which we postulated to maybe be related to the Ashtekar variables of the background geometry. The quantisation of the KG string and the Virasoro-Klein-Gordon algebra presented an anomaly-free algebra, no restrictions on the background dimensionality and no massless spin-1 or spin-2 excitations at the effective level and no negative-norm states in physical processes, with the usual $G_{\mu \nu} , B_{\mu \nu} , \Phi$ fields as part of the spin-2 spectrum.

We also briefly explored an alternative approach to incorporate LQG quantum geometric effects into String Theory by considering Afshordi's model \cite{Afshordi2014Emergent} with some key insights from LQG complementing said model, creating a more concrete picture of the mechanism behind the emerging dimensions and how ultimately spacetime (and thus the true quantum gravitational field) might be actually $(2+1)$-dimensional at the Planck scale.

Ultimately one cannot be satisfied with the heuristic action of the form \eqref{eqn:ng-area-corrected}. Therefore in \autoref{sec:holographic-strings} we attempt to build, from the ground up, a modified string action which incorporates the following chain of thought. We seek to describe the behavior of a string in an arbitrarily curved background geometry at a \emph{classical} level. For this an action of the form \eqref{eqn:polyakov-curved} is unsuitable because that action treats the bulk metric as non-dynamical variables. Connection variables are used in LQG to describe the geometry of a manifold. The coupling of the string worldsheet to the bulk geometry should reflect this fact. This leads us to an action of the form \eqref{eqn:poincare-polyakov}. This action no longer possess the global Poincaré symmetry of the original Polyakov action. However it is invariant under \emph{local} Poincaré gauge transformations.

For an arbitrarily (or even a minor deviation from flat space) curved bulk geometry, global Poincaré symmetry of embedding fields is not meaningful. Local Poincaré symmetry, on the other hand, provides a perfectly adequate framework \cite{Hehl2023Four} for constructing a theory of gravity on an arbitrary bulk. Thus we conjecture that the bulk theory is that of gauged Poincaré gravity.

Finally, the interpretation of embedding fields as co-ordinate fields on the bulk and \emph{also} as vectors living in the fundamental representation of the Poincaré group necessitates the introduction of an ``auxiliary'' co-ordinate which has the natural interpretation as a holographic or scale direction.

\subsection{Poincaré Spin Networks}\label{sec:poincare-networks}

The above considerations feed back into our understanding of the bulk gravity theory. Recall, that we started with LQG in which the gauge group for the $3+1$ dimensional bulk is $\sltwoc$. However, our considerations regarding viewing the embedding fields as representations of the Poincaré group imply that the bulk is $4+1$ dimensional, when one includes the emergent (spacelike) holographic direction, and therefore the gravitational theory for the full bulk should be gauge theory of the Poincaré group. Now, if we wish to construct the resulting spin-network and spin-foam states for the bulk by following the LQG quantization procedure, what we end up with will not be the usual $ \mf{su}(2) $ ($ \mf{sl}(2,\C)$) spin-networks (spin-foams), but \emph{Poincaré spin-foams} of the kind recently discussed in a work by Altaisky \cite{Altaisky2024Poincare}.

This suggests that it might make greater sense to use Poincaré spin foams rather than Lorentzian spin foams to describe the behaviour of quantum geometry at the Planck scale. Though even with such a change, the fundamental result of LQG - quantization of areas - which underpins this work, will likely not be affected.

\subsection{Closed String Spin-1 States} \label{sec:spin-1-states}

In section \ref{sec:quantum-kg-string} we saw that our area-corrected model allows closed string spin-1 excitations, which is due to our solution not having distinct left and right sectors and the physical states not being invariant under rigid rotations of the string. Closed string excitations are usually related to the gravitational field and background spacetime itself, so what could the vector states of the closed string relate to? Well, in section \ref{sec:gauging-lt} we discussed a possible formulation of ST in terms of a Poincaré connection $\mc{A}_{\mu}$, which is an ingredient in describing the gravitational field in tetrad-connection \cite{Bodendorfer-LQG} formulation, and so this seems to be the natural candidate as to what the first level vector excitation of the closed string could be,
\begin{equation} \label{eqn:connection-string-vector-excitation}
	\mc{A}_{\mu} \sim \eta _{\mu \nu} (\Hat{a}^{\nu} _{1})^{\dagger} \ket{0} .
\end{equation}
Another candidate for what the vector excitations could relate to is the Ashtekar electric field $\widetilde{E}^{\mu} _{I}$,
\begin{equation} \label{eqn:AEF-string-vector-excitation}
	\widetilde{E}^{\mu} _{I} \sim (\Hat{a}^{\mu} _{I})^{\dagger} \ket{0} .
\end{equation}

\begin{acknowledgments}
	DV would like to thank Thomas van Riet for going over an early draft of this work and for his valuable comments. DV and LdS would both like to thank David Schmidtt for his very valuable observations regarding the nature of the $\Delta$ field and its relation to the Kalb-Ramond field, and also the equivalence of the area corrected action with a Born-Infeld model in the worldsheet. Finally, we would like to humbly dedicate this work to the memory of the victims of the genocide in Gaza.
\end{acknowledgments}

\appendix

\section{Strings in Background Fields and Emergent Gravity}\label{sec:strings-gravity}

Let us begin by recalling the Polyakov action for the bosonic string.
\begin{equation}\label{eqn:polyakov-action}
	S_{P} = -\frac{T}{2} \int \diff \tau \wedge \diff \sigma \sqrt{-g} g^{ab} \partial_a X^\mu \partial_b X^\nu \eta_{\mu\nu}.
\end{equation}
Here $\mu, \nu \in \{0,1,\ldots,D-1\}$ are co-ordinates of the $D$ dimensional worldvolume (the geometry in which the string propagates), $a,b \in \{0,1\}$ are co-ordinates on the string worldsheet, $X^\mu$ are the embedding co-ordinates which specify the location of a point on the string worldsheet in the bulk worldvolume, $\eta_{\mu\nu}$ is the flat Minkowski metric on the worldvolume, $g_{ab}$ is the metric on the string worldsheet, $T$ is the string tension and $\tau, \sigma $ are the co-ordinates on the string worldsheet.

Now, one proceeds in the usual way by determining the symmetries of the action \eqref{eqn:polyakov-bimetric}, varying the action to find the equations of motion, fixing the gauge using the Weyl freedom of the worldsheet and then solving the classical equations of motion. Imposition of (bosonic) commutation relations on the operator versions of the embedding fields $\hat X^\mu$ then leads us to the description of the quantum state of the bosonic string in terms of an infinite ladder of harmonic oscillators which obey the Virasoro algebra.

The obvious drawback of this approach is that the background metric is non-dynamical and is fixed to be the flat Minkowski metric $\eta_{\mu\nu}$. Clearly, one would like to be able to understand the physics of a string propagating on an arbitrary curved background. It wouldn't make much sense to refer to string theory as a theory of ``quantum gravity'' if strings can only be described on flat backgrounds. The way this is accomplished is by treating the metric of the worldvolume as a ``background field'' $G_{\mu\nu}$, in terms of which the Polyakov action becomes:
\begin{equation}\label{eqn:polyakov-curved-bulk}
	S'_{P} = -\frac{T}{2} \int \diff \tau \wedge \diff \sigma \sqrt{-g} g^{ab} \partial_a X^\mu \partial_b X^\nu G_{\mu\nu}(X),
\end{equation}
where the bulk metric is now a function of the bulk coordinates $G_{\mu\nu}(X)$. Now this metric is still non-dynamical because the action \eqref{eqn:polyakov-curved-bulk} does not contain any terms with time derivatives of $G_{\mu\nu}$. However, if we view the action purely as a two-dimensional theory of $D$ scalar fields, then the bulk metric can be viewed as a collection of \emph{coupling constants}, rather than a dynamical entity which exists independent of the string. One can now proceed in the usual manner for any field theory and calculate the beta function of the coupling constants of the theory as a function of the energy scale.

Then, as shown long ago by Friedan \cite{Friedan1985Nonlinear} (see also \cite{Callan1989Sigma}, \cite{Callan1985Strings} or \cite[Sec 3.7]{Polchinski1998aString}, \cite[Sec 7.2]{Tong2010Lectures} for a more pedagogical explanation), the graviton beta function is proportional to $R_{\mu\nu}$, the Ricci curvature of the bulk geometry. The requirement of Weyl invariance of the string worldsheet implies that this beta function should vanish:
\begin{equation}\label{eqn:ricci-flat}
	\beta(G_{\mu\nu}) \propto R_{\mu\nu} = 0.
\end{equation}
Therefore we find that conformal invariance of the string worldsheet \emph{implies} that the background geometry in which the string propagates satisfies the \emph{vacuum} Einstein equations. It is this result which is often cited as evidence for the claim that string theory is a \emph{background independent} theory of quantum gravity.

\section{Two Expressions for Area}\label{eqn:area-formulae}

There is another explanation for why it makes sense to promote the $X$'s to local fields. This can be found in terms of the expression for the area of a 2-surface. We have a manifold $\mc M$, within which is embedded a two dimensional submanifold $\mc N$. Let $(\tau,\sigma)$ be the location of a point on the worldsheet and $X^\mu(\tau, \sigma)$ be the location of that same point in terms of co-ordinates on $\mc M$. Then we can define the induced metric on $\mc N$, $h_{ab}$ in terms of derivatives of the embedding fields (the pullback):

\begin{equation}\label{eqn:ws-metric}
	h_{ab} = \frac{\partial X^\mu}{\partial x^a} \frac{\partial X^\nu}{\partial x^b} G_{\mu\nu},
\end{equation}
where $G_{\mu\nu}$ is the metric in the ambient space $\mc M$. Now, we can calculate the area of a small patch of $\mc P \in \mc N$ by taking the square root of the determinant of the induced metric $\sqrt{- \det(h)}$ and integrating over that patch:
\begin{equation}\label{eqn:patch-area-v1}
	A^{(1)}_{\mc P} = \int_{\mc P} d^2 x \, \sqrt{- \det(h)}.
\end{equation}
This is, of course, the expression for the Nambu-Goto action - modulo some constants - which we all know and love. In LQG we don't have a notion of embedding fields. Since we are dealing with gauge invariant observables constructed from holonomies of a connection along a 1d-curve and the flux of a tetrad across a 2-surface, the embedding coordinates of the curves and surface turn to not be needed. When promoting classical observables into quantum operators acting on lines and surfaces the embedding coordinates of the lines and surfaces are not relevant as long as we work with gauge invariant quantities.

How then does one define the area of a surface in LQG? It is defined as the magnitude of the area two form. Recall that given a tetradic basis $e_\mu^I$ for a local Lorentz frame whose metric is given by $g_{\mu\nu}$, we can write the metric in terms of the tetrad in the usual manner:
\begin{equation}\label{eqn:metric-from-tetrad}
	g_{\mu\nu} = e_\mu^I e_\nu^J \eta_{IJ},
\end{equation}
where $\eta_{IJ}$ is the flat Minkowski metric on the ``internal'' flat space. If we consider any two dimensional surface in the local Lorentz frame, then its area can be written in terms of the vielbein as:
\begin{equation}
	A^{(2)}_{\mc P} = \int_{\mc P} \Tr (e \wedge e),
\end{equation}
where $e = e_{\mu} \diff x^{\mu}$ and the trace is taken over the ``internal'' Lorentz indices.

Now, if we compare the two expressions \eqref{eqn:ws-metric} and \eqref{eqn:metric-from-tetrad} we notice an obvious parallel between the derivatives $\partial X^\mu/\partial x^a$ of the embedding co-ordinates and the vielbein $e_\mu^I$. This similarity becomes more apparent once we write down the expression for the area of a 2-surface by taking the determinant in \eqref{eqn:patch-area-v1} and comparing that to the expression obtained using frame fields.

\textbf{From Embedding Fields:}

We take the bulk metric $g_{\mu\nu}$ to be flat. Let $\{\tau,\sigma\} := \{x^0, x^1\}$ be the spacetime co-ordinates on the worldsheet. Derivatives of the $X$'s in terms of these co-ordinates can be written as:
\begin{equation}
	\dot X := \left\{\frac{\partial X^\mu}{\partial \tau}\right\}; \quad X' := \left\{\frac{\partial X^\mu}{\partial \sigma}\right\}
\end{equation}
we obtain the following form for the determinant of the worldsheet metric (\eqref{eqn:patch-area-v1}) in terms of the embedding fields:
\begin{align}\label{eqn:area-embedding}
	\det h & = \begin{vmatrix}
		           h_{00} &  & h_{01} \\
		           h_{10} &  & h_{11}
	           \end{vmatrix} = \begin{vmatrix}
		                           \dot X^2        &  & \dot X \cdot X' \\
		                           \dot X \cdot X' &  & X'^2
	                           \end{vmatrix} \nonumber \\
	       & = \dot X^2 X'^2 - (\dot X \cdot X')^2
\end{align}

\textbf{From Vielbein:}

We can proceed to obtain the area of a 2-surface from the bulk vielbein as follows. Starting with the bulk vielbein $e_\mu^I$, we perform a local Lorentz transformation such that two ``legs'' of the vielbein are tangent to the worldsheet surface. We then project down the vielbein $e_\mu^I$ to a dyad $e_a^I$ living in the tangent space of the worldsheet, in terms of which the worldsheet metric can be written as:
\begin{equation}
	h_{ab} = e_a^I e_b^J \eta_{I J},
\end{equation}
where $\eta_{I J}$ is the flat Minkowski metric in group space. The determinant then becomes:
\begin{align}\label{eqn:area-vielbein}
	\det h & = \begin{vmatrix}
		           h_{00} &  & h_{01} \\
		           h_{10} &  & h_{11}
	           \end{vmatrix} = \begin{vmatrix}
		                           e_t^I e_t^J & e_t^I e_x^J \\
		                           e_x^I e_t^J & e_x^I e_x^J
	                           \end{vmatrix} \eta_{I J} \nonumber       \\
	       & = \vec e_t{}^2 \vec e_x{}^2 - (\vec e_t \cdot \vec e_x)^2,
\end{align}
where $\vec e_t \cdot \vec e_x = e_t^I e_x^J \eta_{I J}$. It is clear by comparing the two expressions \eqref{eqn:area-embedding} and \eqref{eqn:area-vielbein} that we have the following correspondence between the vielbein and the $X$ derivatives:
\begin{equation}
	\vec e_\tau \equiv \frac{\partial \vec X}{\partial \tau}; \quad \vec e_\sigma \equiv \frac{\partial \vec X}{\partial \sigma}.
\end{equation}
In other words, we can directly identify the zweibein with the derivatives of the embedding fields. Of course, this is not altogether surprising. If we consider any manifold on which we have some set of fields $\{X^I(x^\mu)\}$, where $I$ is valued in the Lie algebra $\mf{g}$ of some group $\mc G$, as a function of some  ``fiducial'' co-ordinates $\{x^\mu\}$, then we can always define a set of $\mf{g}$-valued vector fields:
$$
	e^I_\mu := \frac{\partial X^I}{\partial x^\mu}.
$$
In the usual formulation of the string action, the embedding fields are \emph{not} Lie-algebra valued but are c-ordinate fields on what is generally taken to be a flat background. For this correspondence between vielbein and pullback Jacobian to work we have to view the $X$ fields as taking values in a Lie algebra. But these same $X$ fields \emph{also} serve as local coordinates for the bulk. If we want the coordinates to change from point to point in the bulk geometry, we must therefore require that the connection $\mc A$ in \eqref{eqn:covariant-deriv-v2} can never vanish completely! If $\mc A$ vanishes then that will imply that the fields $X$ don't change from one point to the next and cannot therefore be viewed as co-ordinate functions. Thus the amplitude $|\mc A|$ can be arbitrarily small but it can never be zero.

\section{Modified Nambu Goto Equations of Motion}\label{sec:modified-eom}

Following the procedure done in \cite{Zwiebach2009A-First}, the correction (\ref{eqn:modified-ng-action}) does not change the form of the equations of motion and neither of the boundary conditions, i.e they are still given by
\begin{equation} \label{eqn:modified-ng-eom}
	\mathrm{EoM:} \partial _{\tau} \PS ^{\tau} _{\mu} + \partial _{\sigma} \PS ^{\sigma} _{\mu} = 0
\end{equation}
and
\begin{equation} \label{eqn:modified-ng-bc}
	\mathrm{B.C.:} \PS ^{\sigma} _{\mu} \delta X^{\mu} \big| _{\sigma = 0} ^{\sigma = \sigma _1} = 0 ,
\end{equation}
where
\begin{align} \label{eqn:modified-ng-tau-currents}
	\PS ^{\tau} _{\mu} & = \frac{\partial \Lagr ' _{N G}}{\partial \Dot{X}^{\mu}} \nonumber                                                                 \\
	                   & = - T \frac{(\Dot{X} \cdot X') X' _{\mu} - (X')^2 \Dot{X}_{\mu}}{\sqrt{- h}} \left( 1 + \frac{g \Delta}{2 (- h)} \right) \nonumber \\
	                   & = \PS ^{\tau} _{\mu (N G)} \left( 1 + \frac{g \Delta}{2 (- h)} \right)
\end{align}
\begin{align} \label{eqn:modified-ng-sigma-currents}
	\PS ^{\sigma} _{\mu} & = \frac{\partial \Lagr ' _{N G}}{\partial X^{\prime \mu}} \nonumber                                                                     \\
	                     & = - T \frac{(\Dot{X} \cdot X') \Dot{X}_{\mu} - (\Dot{X})^2 X' _{\mu}}{\sqrt{- h}} \left( 1 + \frac{g \Delta}{2 (- h)} \right) \nonumber \\
	                     & = \PS ^{\sigma} _{\mu (N G)} \left( 1 + \frac{g \Delta}{2 (- h)} \right)
\end{align}
and $\PS ^{\tau} _{\mu (N G)}$, $\PS ^{\sigma} _{\mu (N G)}$ are the regular Nambu-Goto world-sheet currents.

In the following, we parametrize the string WS using the static gauge $\tau = t$ and transverse gauge
\begin{equation} \label{eqn:transverse-'gauge'}
	\frac{\partial X}{\partial s} \cdot \frac{\partial X}{\partial t} = 0 ,
\end{equation}
where $s$ is the length along the string parameter since this gives the motion of the string in terms of transverse velocity, which is a natural dynamical variable of the string. We find that the new form for the conserved energy per infinitesimal piece of the string is the regular energy of the Nambu-Goto string times the correction factor
\begin{equation} \label{eqn:string-energy}
	E \diff \sigma = T \gamma _{v_{\perp}} \left( 1 + \frac{g \Delta}{2 (- h)} \right) \diff s ,
\end{equation}
and the form of the wave equation maintains its shape as in
\begin{equation} \label{eqn:corrected-ng-wave-equation}
	\mu _{eff} \frac{\partial ^2 \Vec{X}}{\partial t^2} - \frac{\partial}{\partial s} \left[ T_{eff} \frac{\partial \Vec{X}}{\partial s} \right] = 0 ,
\end{equation}
where the effective mass density and effective tension are given by
\begin{align}
	\mu _{eff} & = \mu \gamma _{v_{\perp}} \left( 1 + \frac{g \Delta}{2 (- h)} \right) \label{eqn:effective-mass}            \\
	T_{eff}    & = \frac{T}{\gamma _{v_{\perp}}} \left( 1 + \frac{g \Delta}{2 (- h)} \right) . \label{eqn:effective-tension}
\end{align}

Rewriting eq(\ref{eqn:corrected-ng-wave-equation}) in terms of $\sigma$ derivatives and parameterizing $\sigma$ as (let $F^+ = 1 + g \Delta / 2 (- h)$)
\begin{equation} \label{eqn:sigma-parameterization}
	\sigma (q) = \frac{1}{T} \int _{0} ^{q} \diff E \frac{1}{(F^+)^2} ,
\end{equation}
we find the variable-speed wave equation
\begin{equation} \label{eqn:simplified-modified-wave-eqn}
	(F^+)^2 \frac{\partial ^2 \Vec{X}}{\partial t^2} - \frac{\partial ^2 \Vec{X}}{\partial \sigma ^2} = 0 ,
\end{equation}
from which we can see that the correction factor modifies the speed of propagation of the wave in the string as
\begin{equation} \label{eqn:wave-speed-corrected}
	v = \frac{c}{F^+}.
\end{equation}

\section{Relating Nambu-Goto and Polyakov analysis} \label{sec:polyakov-ng-relation}

At first glance, our analysis of the Polyakov string with inverse metric correction seems to not be related to what we found in the analysis of the Nambu-Goto string. However, as pointed out in \cite{Grimshaw2010Homogenization}, wave equations with a variable speed (the result of the Nambu-Goto analysis) can be turned into Klein-Gordon equations (what we found in section \ref{sec:inverse-metric-Polyakov}). While their work deals only with space-varying wave speed, it is straightforward to generalise the procedure to also include time dependence.

Following the procedure found in \cite{Grimshaw2010Homogenization}, we start with eq (\ref{eqn:simplified-modified-wave-eqn}) obtained in section \ref{sec:modified-eom}, renaming $\sigma$ as $x$ and dividing by $(F^{+})^2$ such that we have $v^2$ on the second term like the mentioned work, and perform a change of variables $(t , x) \rightarrow (\tau (t , x) , \sigma (t , x))$, where without loss of generality (one needs just to rescale and/or rotate the coordinates) we set
\begin{multline} \label{eqn:new-coordinates-constraints-quadratic}
	\ \ \ \ \ \ \ \ \ \ (\partial _{t} \sigma)^2 - v^2 (\partial _{x} \sigma)^2 = v^2 (\partial _{x} \tau)^2 - (\partial _{t} \tau)^2 \\
	\partial _{t} \tau \partial _{t} \sigma = v^2 \partial _{x} \tau \partial _{x} \sigma , \ \ \ \ \ \ \ \ \ \ \ \
\end{multline}
or more compactly,
\begin{multline} \label{eqn:new-coordinates-constraints-compact}
	\ \ \ \ \ \ \ \ \ \ \ \ \ \ \ \ \ \ \ \ \ \ \ \ \ \ \ \partial _{t} \tau = v \partial _{x} \sigma \\
	\partial _{t} \sigma = v \partial _{x} \tau \ \ \ \ \ \ \ \ \ \ \ \ \ \ \ \ \ \ \ \ ,
\end{multline}
which is equivalent to
\begin{multline} \label{eqn:new-coordinates-constraints-second-order}
	\ \ \ \ \ \ \ \ \ \ \ \partial _{t} ^2 \tau - v^2 \partial _{x} ^2 \tau = \partial _{t} v \partial _{x} \sigma + v \partial _{x} v \partial _{x} \tau \\
	\partial _{t} ^2 \sigma - v^2 \partial _{x} ^2 \sigma = \partial _{t} v \partial _{x} \tau + v \partial _{x} v \partial _{x} \sigma . \ \ \ \ \ \ \
\end{multline}
This turns eq (\ref{eqn:simplified-modified-wave-eqn}) into
\begin{multline} \label{eqn:modified-wave-eqn-new}
	\partial _{\tau} ^2 X^{\mu} - \partial _{\sigma} ^2 X^{\mu} + \\
	+ \frac{\partial _{t} v \partial _{x} \sigma + v \partial _{x} v \partial _{x} \tau}{((\partial _{t} \tau)^2 - v^2 (\partial _{x} \tau)^2)} \partial _{\tau} X^{\mu} + \\
	+ \frac{\partial _{t} v \partial _{x} \tau + v \partial _{x} v \partial _{x} \sigma}{((\partial _{t} \tau)^2 - v^2 (\partial _{x} \tau)^2)} \partial _{\sigma} X^{\mu} = 0 .
\end{multline}
Next, we introduce $X^{\mu} = \kappa (\tau , \sigma) W^{\mu} (\tau , \sigma)$ such that terms proportional to first derivatives of $W^{\mu}$ cancel out. This is obtained by choosing $\kappa = v^{1/2}$ and concluding that $v$ only depends on $x$. With this, eq (\ref{eqn:modified-wave-eqn-new}) turns into a Klein-Gordon equation
\begin{equation}
	\partial _{\tau} ^2 W^{\mu} - \partial _{\sigma} ^2 W^{\mu} + \mu ^2 W^{\mu} = 0 ,
\end{equation}
with \textquotedblleft mass" squared given by
\begin{equation}
	\mu ^2 = \frac{(\partial _{x} v)^2 - 2 v \partial _{x} ^2 v}{4 \left( (\partial _{t} \tau)^2 - v^2 (\partial _{x} \tau)^2 \right)}
\end{equation}
and $\tau (t , x)$ satisfying constraints (\ref{eqn:new-coordinates-constraints-compact}). This shows that our analysis of the action (\ref{eqn:inverse-metric-Polyakov}) is indeed in agreement with the analysis of the inverse area corrected Nambu-Goto action (\ref{eqn:modified-ng-action}).

\bibliographystyle{JHEP3}

\bibliography{lqg-strings.bib}

\end{document}